\documentclass[a4paper,fleqn,usenatbib]{mnras}
\usepackage{graphicx}	
\usepackage{amsmath}	
\usepackage{amssymb}	
\usepackage{gensymb}
\usepackage[justification=centering]{caption}
\usepackage{nth}
\usepackage{xcolor}
\usepackage{mathptmx}
\usepackage{txfonts}
\usepackage[none]{hyphenat}
\usepackage[T1]{fontenc}
\usepackage{ae,aecompl}
\usepackage{caption}
\usepackage{graphicx}	
\usepackage{amsmath}	
\usepackage{amssymb}	
\usepackage{lineno}

\title[Brightest Cluster Galaxies in SPIDERS]{Exploring Relations Between BCG \& Cluster Properties in the SPectroscopic IDentification of eROSITA Sources Survey from $0.05 < z < 0.3$}
\author[K. E. Furnell et al.]{
Kate E. Furnell,$^{1}$\thanks{E-mail: K.E.Furnell@2015.ljmu.ac.uk}
Chris A. Collins,$^{1}$
Lee S. Kelvin$^{1},$
Nicolas Clerc$^{2,3},$
Ivan K. Baldry$^{1}$\newauthor
Alexis Finoguenov$^{2,4},$
Ghazaleh Erfanianfar$^{2},$
Johan Comparat$^{2}$\newauthor
\& Donald P. Schneider$^{5,6}$\\
$^{1}$Astrophysics Research Institute, Liverpool John Moores University, IC2, Liverpool Science Park, 146 Brownlow Hill, Liverpool, L3 5RF\\
$^{2}$ High-Energy Astrophysics Group, Max Planck Institute for Extraterrestrial Physics, Giessenbachstrasse, 85741, Garching, Germany\\
$^{3}$ IRAP, Universit\'{e} de Toulouse, CNRS, UPS, CNES, Toulouse, France \\
$^{4}$Department of Physics, University of Helsinki, PL 64 (Gustaf H\"{a}llstr\"{o}min Katu 2), D318, 
Helsinki, Finland\\
$^{5}$ Department of Astronomy and Astrophysics, 
The Pennsylvania State University,
University Park, PA 16802\\
$^{6}$ Institute for Gravitation and the Cosmos, The Pennsylvania State University,
University Park, PA 16802\\
}

\date{Accepted 2018 March 28. Received 2018 March 9; in original form 2017 December 1}

\pubyear{2017}

\begin{document}

\label{firstpage}
\pagerange{\pageref{firstpage}--\pageref{lastpage}}
\maketitle
\begin{abstract}
We present a sample of 329 low to intermediate redshift ($0.05  < z < 0.3$) brightest cluster galaxies (BCGs) in X-ray selected clusters from the SPectroscopic IDentification of eRosita Sources (SPIDERS) survey, a spectroscopic survey within Sloan Digital Sky Survey-IV (SDSS-IV). We define our BCGs by simultaneous consideration of legacy X-ray data from ROSAT, maximum likelihood outputs from an optical cluster-finder algorithm and visual inspection. Using SDSS imaging data, we fit S\'ersic profiles to our BCGs in three bands (\textit{g}, \textit{r}, \textit{i}) with \textsc{SIGMA}, a \textsc{GALFIT}-based software wrapper. We examine the reliability of our fits by running our pipeline on ${\sim}10^{4}$ psf-convolved model profiles injected into 8 random cluster fields; we then use the results of this analysis to create a robust subsample of 198 BCGs. We outline three cluster properties of interest: overall cluster X-ray luminosity ($L_{X}$), cluster richness as estimated by \textsc{redMaPPer} ($ \lambda $) and cluster halo mass ($M_{200}$), which is estimated via velocity dispersion. In general, there are significant correlations with BCG stellar mass between all three environmental properties, but no significant trends arise with either S\'{e}rsic index or effective radius. There is no major environmental dependence on the strength of the relation between effective radius and BCG stellar mass. Stellar mass therefore arises as the most important factor governing BCG morphology. Our results indicate that our sample consists of a large number of relaxed, mature clusters containing broadly homogeneous BCGs up to $z \sim 0.3$, suggesting that there is little evidence for much ongoing structural evolution for BCGs in these systems.
\end{abstract}




\section{Introduction}\label{intro}
Galaxy clusters, the nodes of the cosmic web, are the largest gravitationally bound structures in the Universe, most of which host a massive `brightest cluster galaxy', or BCG, at their core. BCGs are unique objects, thought to represent a distinct population \citep{1987ApJ...313...59D} separate from the general elliptical non-BCG population \citep{2007MNRAS.379..867V} The visual distinctions are often pronounced at local scales, with many BCGs hosting distinct, extended envelopes of stars \citep[e.g.,][]{1965ApJ...142.1364M, 1986ApJS...60..603S}. They also exhibit extraordinary homogeneity, with many studies finding uniformly-luminous BCGs in similarly massive clusters up to redshifts of $z \sim 1$ \citep{1998MNRAS.297..427A, 2009Natur.458..603C, 2008MNRAS.387.1253W}. This property conflicts with current simulations, which struggle to reproduce the observed homogeneity seen in BCGs up to redshifts of $z \sim 1$. Many simulations are successful in reproducing the colours of observed BCGs, but they fail to accurately model BCG growth, over-predicting the observed amount by between a factor of 2$-$4 \citep[e.g.,][]{2007MNRAS.375....2D, 2013MNRAS.436.1750R}. Although improvements have been made in recent years, most simulations still cannot reproduce the observed morphologies of BCGs in general, especially within their central regions \citep[e.g.,][]{2015MNRAS.451.1177L, 2012ApJ...759...43T}. It has been recognised therefore that understanding the growth of BCGs and the global accumulation of baryons in cluster cores may be critical for solving some of the discrepancies suffered by clusters in hydrodynamical simulations, such as the well-documented cuspy halo problem \citep[e.g.,][]{1996ApJ...462..563N}.
\par In contrast to nearby clusters, identifying the BCG is often non-trivial in more distant clusters. It has been established for some time that, for the majority of galaxy clusters which are dynamically `relaxed', BCGs reside close to the peak of cluster X-ray emission \citep[e.g.,][]{2004ApJ...617..879L, 2010A&A...513A..37H}. This X-ray peak signifies inflow of the intracluster medium (ICM), the hot, diffuse gas between clusters, into the cluster core; this is indicative of a deep potential well \citep[e.g.,][]{1973ApJ...184L.105L, 1977MNRAS.180..479F}. However, for clusters out of dynamic equilibrium (i.e. ones which have undergone recent mergers with neighbouring groups), this technique is ineffective, with multiple, similarly-luminous candidate BCGs and X-ray peaks often being present \citep[e.g.,][]{2002ApJ...567L..27M}. Merger activity is more common in high-redshift systems, leading many to conclude that heirarchical merging is the dominant mechanism behind the growth of galaxy clusters \citep[e.g.,][]{1991ApJ...379...52W}. Nevertheless, there are many issues to consider when quantifying cluster parameters such as mass, with numerous proxies used throughout the literature with various caveats \citep[e.g., caustics, velocity dispersions, richnesses, X-ray mass-temperature scaling and weak gravitational lensing; see][]{2015MNRAS.449.1897O}.
\par Due to the location of BCGs at the peak of X-ray emission in relaxed clusters, there is speculation that BCGs predominantly grow through ongoing star formation as a result of condensing cluster gas regulated by some feedback mechanism \citep{1994MNRAS.267..779F, 2014ApJ...785...44M, 2016arXiv160702212V}. For many hydrodynamical simulations, AGN activity remains the favoured dominant candidate for feedback \citep[e.g.,][]{2016arXiv160302702M, 2015MNRAS.446..521S, 2014MNRAS.444.1518V}, although some studies are moving towards closer examination of external baryonic processes such as ram-pressure stripping and shock heating \citep[e.g.,][]{2016A&A...591A..51S}. Indeed, high rates of star formation ($10^{1}-10^{2} \mathrm{ \ M_{\odot}yr^{-1}}$), have been detected in some BCGs within clusters hosting cool-cores \citep[e.g.,][]{2008ApJ...681.1035O, 2001MNRAS.328..762E} as well as enhanced AGN activity \citep[e.g.,][]{1990AJ.....99...14B} and giant molecular gas outflows \citep[e.g.,][]{2014ApJ...784...78R}. However, mass deposition rates may still be too slow to reproduce the mass range of BCGs through in-situ star formation alone \citep[e.g.,][]{2006PhR...427....1P}, alongside strong cool-core systems being relatively rare at higher redshifts where observations suggest BCGs gain the bulk of their mass (\citealt{2007hvcg.conf...48V}; \citealt{2009Natur.458..603C}). 
\par \par BCG formation scenarios based on classical hierarchical merging have risen as a popular alternative to growth through star formation since $z \sim 1$ \citep[e.g.,][]{1975ApJ...202L.113O,1985ApJ...289...18M}. Broadly, these models fall into two categories: `galactic merging', where many similarly-sized galaxies violently merge together in a short amount of time to form the BCG, or `galactic cannibalism', a slower process where mass is accumulated over time from smaller galaxies sinking to the bottom of the cluster potential well. Currently, a two-step scenario behind BCG formation is favoured \citep[e.g., ][]{2012ApJ...754..115J}, with the bulge forming first at early times ($z > 2$) followed by the envelope feature at late times ($z < 1$). This approach has gained popularity as an explanation behind the `cD-type' morphologies (i.e. bulge + halo) seen in many BCGs at low redshift, as well as the apparent `puffing up' of elliptical-types \citep[e.g. ][]{2008ApJ...677L...5V, 2008ApJ...688...48V}. Indeed, there is convincing indirect evidence that merger events happen at some point in the formation history of BCGs, with many examples of BCGs hosting multiple nuclei at their centres \citep[e.g.,][]{1988ApJ...325...49L, 1983ApJ...264..337S}
\par The purpose of this work is to investigate a sample of BCGs in order to examine their properties with respect to the properties of their host clusters. As aforementioned, many studies have found links of varying strength between the properties of BCGs and their host clusters up to redshifts of $z \sim 1$ \citep[e.g.][]{2015MNRAS.453.4444Z, 2007MNRAS.379..867V, 2008MNRAS.384.1502S, 2011MNRAS.414..445S}. Numerous studies have reported a positive correlation between BCG stellar masses and environmental properties such as the overall halo mass (e.g. \citealt{1985MNRAS.213..857B}; \citealt{2004ApJ...617..879L} and numerous others), with some finding tentative environmental links between various BCG morphological properties and their host clusters (e.g. \citealt{2005MNRAS.364.1354B}; \citealt{2009MNRAS.398.1129G}). However, others have found little-to-no dependence at low redshift (\citealt{2015MNRAS.453.4444Z}), or have argued that the effect strengthens for central galaxies with late-type morphologies but is not strongly observed in early types (e.g. \citealt{2009MNRAS.394.1213W}). 
\par In this study, we aim to analyse the light profiles of a sample of BCGs with respect to the global cluster environment. We select three major cluster properties to analyse: dynamically-derived halo mass $M_{200}$, X-ray luminosity $L_{X}$ and cluster richness $\lambda$, an estimate of overall cluster membership. This paper is structured as follows: in section \ref{sampleselection}, we discuss the data used in procuring the sample of BCGs used here. In Section \ref{dynamics}, we discuss our analysis of the host cluster dynamics in our sample and our estimation of halo mass. In sections \ref{sigmadescript}, \ref{sigpipedescript} and \ref{simulations}, we discuss the 2D profile fitting techniques used for our sample, as well as a suite of simulations used to test their reliability and bring to light any potential biases. In Section \ref{results}, we describe our results, including a discussion of our method of estimating stellar masses for our objects. We assume a standard  $\mathrm{\Lambda}$CDM concordance cosmology throughout, with $\mathrm{H_{0}} = 70 \ \mathrm{km \ s^{-1}Mpc^{-1}}$, $\mathrm{h_{100} \ = \ 0.7}$, $\mathrm{{\Omega}_{\Lambda}} = 0.7$ and $\mathrm{{\Omega}_{M}} = 0.3$.
\section{Data} 
\subsection{Data Description}
\begin{figure*}
\centering 
\includegraphics[width=17.5cm,height=17.5cm,keepaspectratio,trim={2.5cm 3.5cm 2cm 2cm},clip]{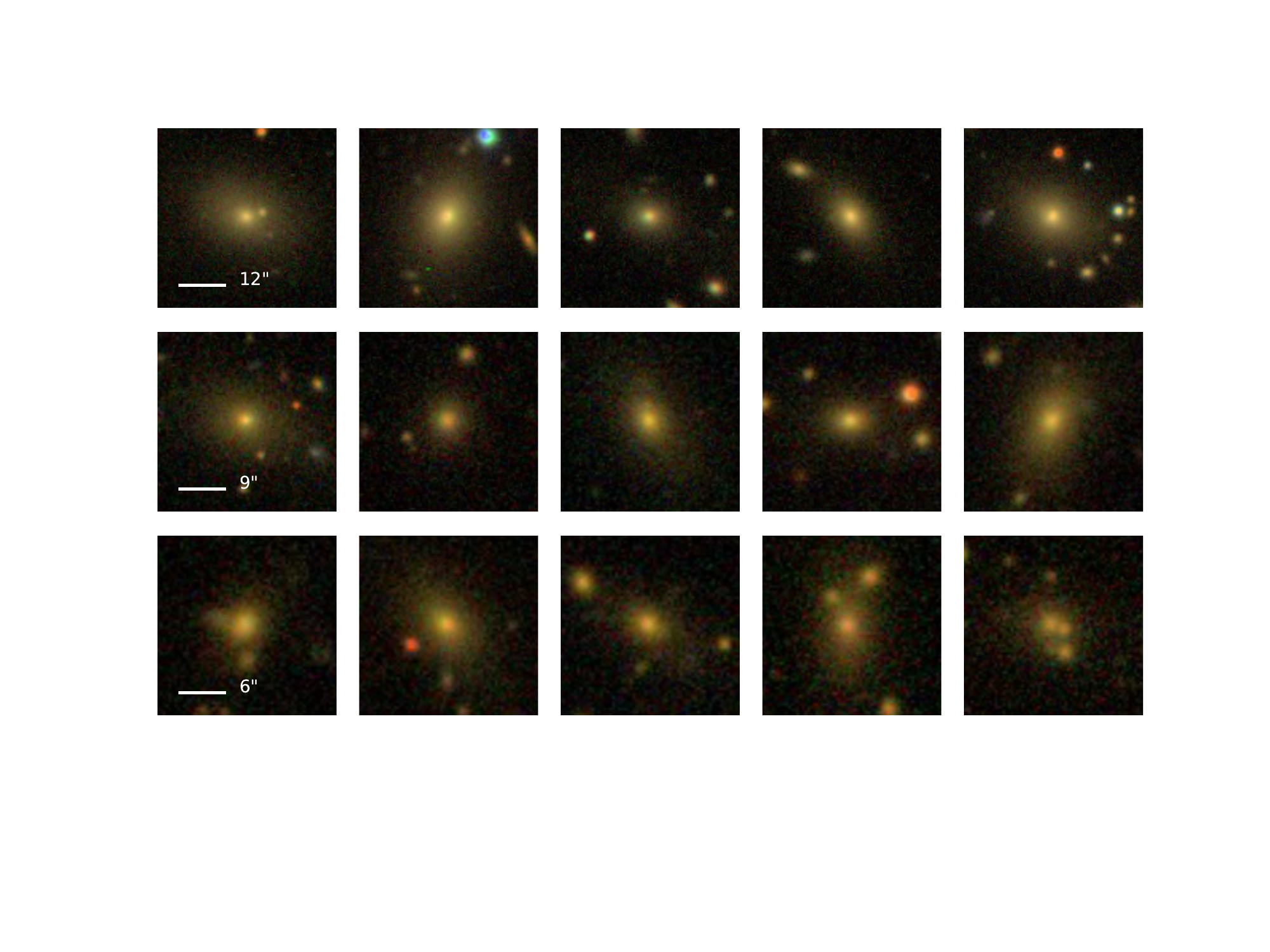}
\caption{Example false-colour SDSS $gri$ composites of 15 BCGs used in this study (100 $\times$ 100 kpc for each tile). Each of the three horizontal panels represents a sample of five randomly drawn BCGs about the $16^{\mathrm{th}}$, $50^{\mathrm{th}}$ and $84^{\mathrm{th}}$ redshift percentiles from left to right (0.1, 0.18 and 0.25,  respectively). The white line represents an equivalent scale of $\sim 25$ kpc (angular equivalent rounded to the nearest arcsecond).}
\label{fig:bcgexamples}
\end{figure*}
This work makes use of observations primarily from two separate sources: X-ray observations from the ROSAT All-Sky Survey (RASS), a shallow, low signal-to-noise ration, all-sky survey of soft X-ray sources running between 1990$-$1999 \citep{1999A&A...349..389V} and optical data from the Sloan Digital Sky Survey \citep[SDSS;][]{2000AJ....120.1579Y}. The spectroscopic data for this work originates from the SDSS-IV (\citealt{2017AJ....154...28B}) SPectroscopic IDentification of eRosita Sources (SPIDERS) survey, which aims to investigate $\sim{10^{4}}$ X-ray targets detectable by eROSITA prior to launch in 2018. eROSITA has been designed with full sky survey capabilities, with high angular resolution (15$"$) and has been predicted to detect all clusters in the Universe above $10^{14} \ \mathrm{M_{\odot}}$ ($\sim 57,500$ clusters out of $\sim 137,000$ total group/cluster detections, \citealt{2012MNRAS.422...44P}), creating the largest and most complete X-ray selected sample of galaxy clusters to date. Optical spectroscopy for SPIDERS are being performed within the eBOSS footprint with the BOSS spectrograph on the SDSS 2.5m Apache Point Observatory telescope \citep[e.g.,][]{2016AJ....151...44D, 2013AJ....146...32S, 2006AJ....131.2332G} and will continue until the end of the project in 2020. 
\par The purpose of the clusters programme in SPIDERS (\citealt{2016MNRAS.463.4490C}) is to obtain observations on a sample of X-ray selected cluster candidates, in order to procure precise redshift measurements of member galaxies. The subsample of cluster candidates in SPIDERS used for this study originates from the COnstrain Dark Energy with X-ray Clusters (CODEX; Finoguenov et al. 2018) survey. Essentially, the CODEX survey combines [$0.5-2$] keV band ROSAT X-ray data with optical data from \textsc{redMaPPer} \citep[i.e. the red-sequence Matched-filter Probabilistic Percolation algorithm; see][]{2014ApJ...785..104R}, an optical cluster finder search on SDSS DR8 data (e.g., \citealt{2011ApJS..193...29A}). In basic terms, the algorithm works through the use of a preliminary `seed' catalogue of red galaxies with spectroscopic data. These `seeds' are used to `train' the algorithm, providing a baseline for photometric redshift estimates. Cluster galaxies are assigned a probability, $P_{MEM}$, of being a member of a detected overdensity of photometrically-grouped red galaxies about a given `seed' galaxy, with the sample undergoing simultaneous filtration in luminosity ($L \geq 0.2L_{\star}$) and projected distance corresponding to the size of a typical cluster at the redshift of the `seed' galaxy. This approach has been enormously successful in both the SDSS and the Dark Energy Survey (DES; \citealt{2005astro.ph.10346T}, see also \citealt{2016ApJS..224....1R}), finding $>10^{5}$ cluster candidates in total. It differs from other commonly-available cluster catalogues, in that it is a `bottom-up' photometric selection with the goal of searching for a red sequence, rather than applying a `friends-of-friends' method to available spectroscopic data; a method more effectively used by spectroscopic surveys with large volumes and high completeness (e.g., GAMA; \citealt{2011MNRAS.416.2640R}).
\par As of November 2015, the CODEX catalogue contains 6693 X-ray detections with corresponding \textsc{redMaPPer} targets within the largest region of the SDSS footprint (see figure 2 of \citealt{2016MNRAS.463.4490C}). The catalogue includes numerous cluster physical properties derived from X-rays or indirectly through X-ray scaling relations, including: rest-frame [$0.5-2$] keV band $L_{X}$ measurements, cluster mass estimates and $T_{X}$ measurements (refer to \citealt{2016arXiv161206871C} for a recent use of the CODEX sample). As of August 2016, 1633 cluster candidates from CODEX with X-ray peaks within 30$'$ of the corresponding luminosity-weighted centroids in SPIDERS have thus far undergone observations (with 1337 having been fully completed). This sample of cluster candidates forms the basis of our BCG sample. All of the optical imaging data used here were taken from the SDSS DR12 release (\citealt{2016AJ....151...44D}; data is calibrated as in DR10), which includes all BOSS data from SDSS-III and offers improved sky-subtraction over previous releases \citep[see][]{2013MNRAS.430..330H, 2011AJ....142...31B}. 
\subsection{Sample Selection}\label{sampleselection}
\par Initial cuts to the SPIDERS sample are outlined in \cite{2016MNRAS.463.4490C}. These include: a SDSS $i$-band magnitude cut (measured in a 2$"$ aperture) of $17.0 \leq \ \mathrm{FIBER2MAG\_I} \ \leq 21.2$ to maximise the efficiency of the redshift detection algorithm and a requirement that any source must have $\geq$ 4 X-ray counts from ROSAT. These cuts are deliberately lax to retain as many candidates as possible; visual inspection efforts on the velocity distributions of cluster members are ongoing by the SPIDERS collaboration at the time of this publication in order to better characterise and identify cluster membership (see upcoming paper, Clerc et al. 2018, see also section \ref{dynamics}).  
\par We impose several extra cuts on the sample of 1633 cluster candidates prior to use. First, we follow the work of \cite{2012MNRAS.422.2213S} and impose a cut on the \textsc{redMaPPer} photometric redshift estimates of our objects as such that $\mathrm{Z\_LAMBDA} \leq 0.3$, in order to ensure decent-quality optical photometry (765 objects). We then adopt a richness cut of LAMBDA\_CHISQ\_OPT $\geq$ 20 to minimise the number of objects in our sample that are not true clusters (Alexis Finoguenov, private communication; see also \citealt{2014ApJ...785..104R}), such as objects in close projection with wildly differing spectroscopic redshifts (see Clerc et al. 2018), or X-ray loud AGN affiliated to a red-sequence identified by \textsc{redMaPPer} (470 objects). The latter detections are accommodated for in richness measurements by applying an AGN HOD model; roughly 2\% of of clusters at $z$ < 0.3 and with richness > 30 are thought to be affected. A final cut required NHASZ $\geq$ 10 (i.e., more than 10 \textsc{redMaPPer}-determined cluster members with at least one SDSS-DR14 spectroscopic measurement spectroscopic observations), following \cite{1995MNRAS.274.1071C}. This selection criterion ensured there were enough remaining members to compute robust velocity dispersion estimates, our chosen method to estimate cluster dynamical masses (see section \ref{dynamics}). After these criteria were applied, a total of 433 cluster candidates remained for visual inspection.
\subsection{BCG Identification}
In this work, an object which we call a `BCG' may not always be the object designated the `brightest' magnitude in a given red sequence detected by \textsc(redMaPPer). Here, we adopt the definition of a `Brightest Cluster Galaxy' as being the `brightest' galaxy in closest angular proximity to the measured X-ray centre of our clusters (e.g. \citealt{2004ApJ...617..879L}). This definition has led several authors to preferentially adopt the term `central galaxy', or CEN/CG (e.g. \citealt{2009MNRAS.398.1129G}; \citealt{2017AJ....153...89O} and others) to clarify their selection, for example, in the case of clusters with comparably bright, infalling galaxies. In our case, however, as we do not probe the group regime where designating a central galaxy is often much more ambiguous (e.g. \citealt{2005MNRAS.356.1293Y}), we use the term as a synonym for the classical, high-mass galaxies which reside in cluster centres.
\par There are several issues which can result in incomplete cluster membership. From an instrumental standpoint, the fibres on the BOSS spectrograph have a minimum separation limit of 62$"$ (55$"$ for the SDSS I/II spectrograph), corresponding to a physical scale of $\sim{100}$ kpc at $z = 0.1$, which may exceed the projected separation between objects in dense source fields (e.g. rich clusters or regions close to the galactic plane), thereby causing incompleteness issues (e.g., \citealt{2007MNRAS.379..867V}; \citealt{2005AJ....130..968M}, which reported a cluster core completeness level of $\sim{50}$\%). In these cases of high source density, the fibre assignment algorithms in the SDSS may break down and measure objects outside of the prior targeting order (e.g. \citealt{2003AJ....125.2276B}). With these caveats in mind, some groups opt to use IFU observations on clusters rather than traditional slit spectroscopy, where simultaneous observations of objects can be taken at increased source density (e.g. \citealt{2015A&A...574A..11K}). Alongside instrumental issues, the survey design, in particular the magnitude threshold of $m_{i}$ > 17.0, may also lead to omitting BCGs in some SPIDERS clusters from the targeting process, due to their brightness in comparison to other cluster members. There are also additional factors which can affect how \textsc{redMaPPer} selects red-sequence objects which it identifies as clusters. For example, the wings of bright sources (i.e., stars) may affect the photometry of fainter objects in close projection on the sky, which can lead to misclassification or omission from the colour-magnitude decision tree method used by \textsc{redMaPPer}. BCGs which have colours statistically atypical with respect to other cluster members may also not be detected, such as those that have high central star formation rates and therefore significantly bluer central regions (e.g. NGC 1275, \citealt{2003MNRAS.344L..43F}). 
\begin{table}
\centering
\caption{Summary of the BCG selection.}
\begin{tabular}{rc|c}
\noalign{\smallskip} \noalign{\smallskip
}
Description & \hfill Number \\
\hline
CODEX-SPIDERS 30$'$ match        & \hfill 1633  	\\
+ Z\_LAMBDA $\leq$ 0.3      & \hfill 765	\\
+ LAMBDA\_CHISQ\_OPT $\geq$ 20        & \hfill 470	\\
+ NHASZ $\geq$ 10        & \hfill \textbf{433}	\\
\hline
Omitted & \hfill \textbf{104} \\
a) Image issues       & \hfill 36	\\
b) Missing       & \hfill 61		\\
c) Major merger        & \hfill 7	\\
\hline
BCG candidates (including legacy SDSS)        & \hfill \textbf{329}	\\

\noalign{\smallskip} \noalign{\smallskip}
\end{tabular}
\label{t:cuts}
\end{table}
\par For these reasons, we deemed it necessary to visually inspect all BCG assignments and clusters by eye. At this stage, we do not make any additional effort to characterise other members than the BCG; details on the automated clipping procedure and velocity dispersion algorithms can be found in section \ref{dynamics}. As aforementioned, we acknowledge that this sample may be slightly biased against objects significantly outside of the red-sequence of their parent cluster because of \textsc{redMaPPer} selection criteria. However, BCGs with abnormally large central star formation rates (> 100 $\mathrm{M_{\odot}} \ \mathrm{yr^{-1}}$) rare with respect to the general population of BCGs at the redshift range in this work \citep[e.g.][]{2009MNRAS.398..133L, 2016MNRAS.461..560G}, with recent estimates on the order of 10\% or lower. Moreover, we believe that any omitted objects are predominantly due to fibre collision problems rather than a result of the \textsc{redMaPPer} algorithm, which we discuss below.
\par The \textit{gri} composite fields were predominantly used to visually inspect the cluster candidates (inspection carried out primarily by K. Furnell). During inspection, member coordinates were extracted from the SPIDERS catalogues and displayed on the images, along with the X-ray centroid, in order to check the robustness of the assignments. We followed a similar selection prescription to \cite{2012MNRAS.422.2213S}, where we select the brightest galaxy at the tip of the red sequence within $R_{200}$ of the X-ray centroid (see Finoguenov et al. in prep). As well as being the most robust identifier of BCGs (e.g. \citealt{2004ApJ...617..879L}), this definition was also found by \cite{2012ApJ...757....2G} to be the best observational proxy for the centre of clusters in general. The $R_{200}$ used here were taken from the CODEX estimates from X-rays.
\par In $\sim{20}$\% of clusters, the true BCG was either incorrect or missing from the SPIDERS catalogues. There were occasional cases where the likelihood ranking of an object being the BCG was incorrectly stated in the catalogue, with the BCG candidate which met our selection criteria ranked lower in the listing (12/433 candidates). More often, however, the BCG listed in the catalogues tended to have been omitted during the SPIDERS selection process (86/433). In these cases, the most visually likely BCG candidate was extracted via \textsc{SExtractor} and cross-matched with all of the spectroscopy available up to DR14 within 5$"$ (over 4 million objects in total). Only 29\% (25/86) of BCGs were recovered in this way ($\sim$ 5\% of the selected sample after cuts). We therefore argue that most omissions are due to fibre collision issues, as there is no spectroscopy for these objects across the SDSS available despite a red sequence identified by \textsc{redMaPPer}. This value is in line with \cite{2014ApJ...785..104R}, which quotes an estimate of misidentified centrals at the 5\% level. Recent efforts have gone into quantifying the ability of the \textsc{redMaPPer} algorithm to correctly centre clusters, with similar results (\citealt{2017arXiv170208614H}). A summary of the various cuts applied to form the sample are shown in table \ref{t:cuts}.

\begin{figure}
\centering 
\includegraphics[width=8.7cm,height=8.7cm,keepaspectratio,trim={2.5cm 0.0cm 2.5cm 0.1cm},clip]{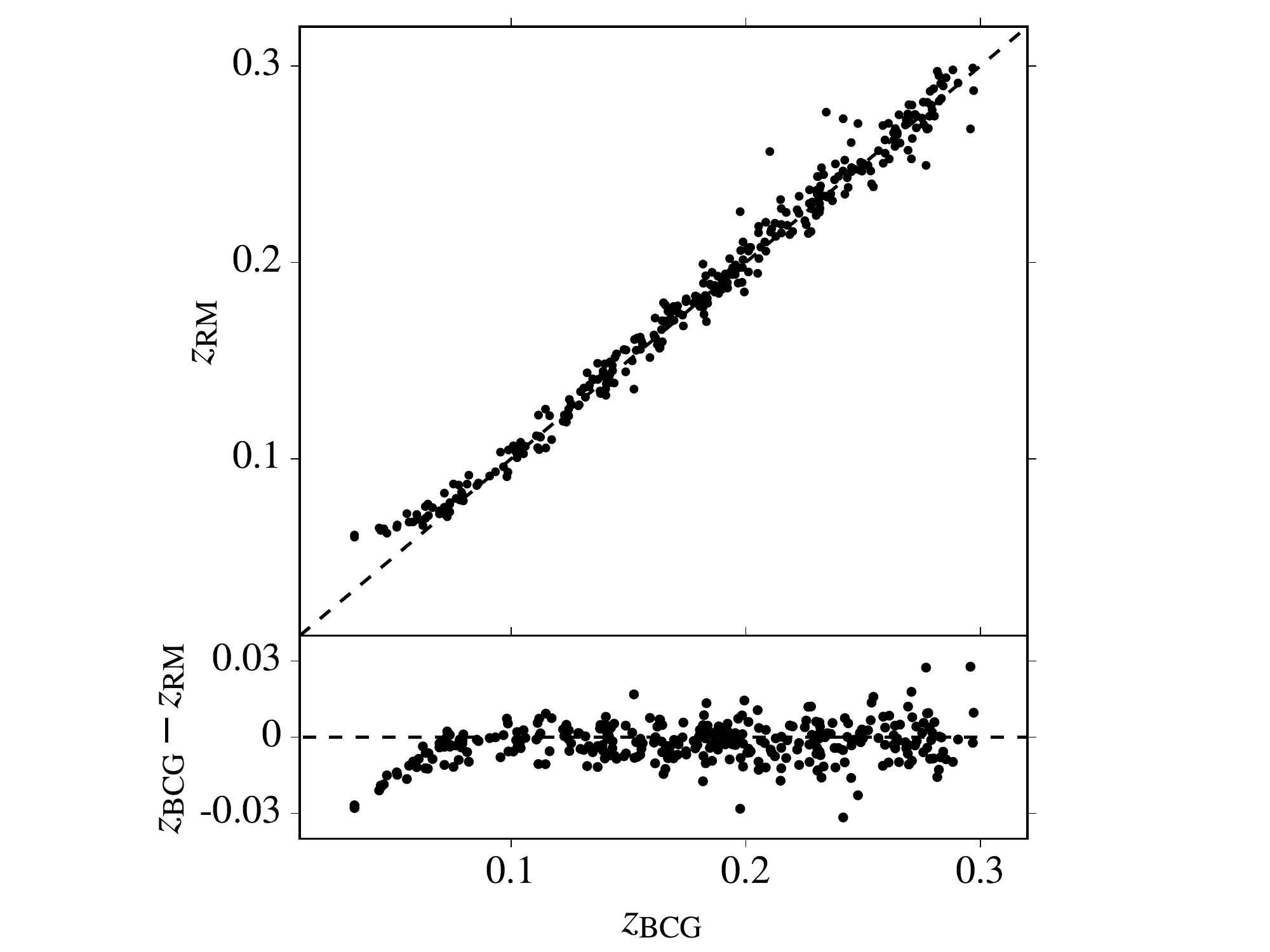}
\caption{Spectroscopic redshift of the visually-inspected BCGs ($z_{\mathrm{BCG}}$) versus photometric cluster redshifit ($z_{\mathrm{RM}}$) comparison for the sample of 329 objects (typical errors are $\Delta z_{\mathrm{BCG}} \sim 10^{-5}$ and $\Delta z_{\lambda} \sim 0.01$ respectively). The black dashed line is the 1:1 relation. In general, the BCG redshifts agree well with the cluster photometric values, albeit with some scatter at higher $z$ (see section \ref{dynamics}). The discrepancies in photometric redshift at $z \leq 0.08$ are discussed in detail in \protect\cite{2014ApJ...785..104R}.}
\label{fig:zphotz}
\end{figure}

\begin{figure}
\centering 
\includegraphics[width=8.0cm,height=8.0cm,keepaspectratio,trim={0cm 0.1cm 0cm 0cm},clip]{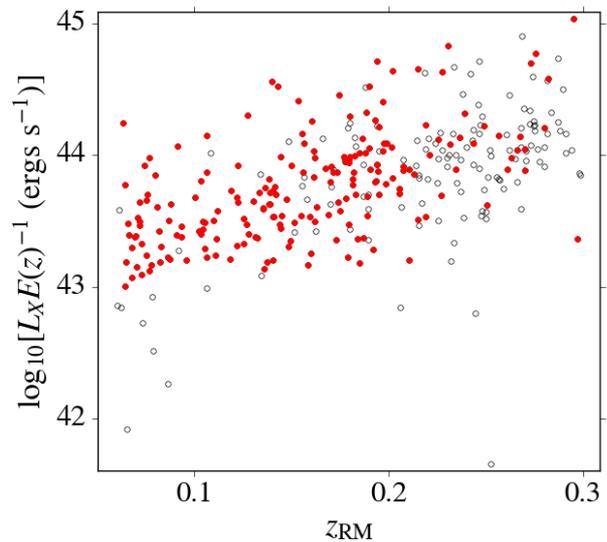} 
\caption{$L_{X}-z_{\mathrm{RM}}$ distribution for the clusters in our sample. The black circles show the initial 329 clusters which passed our visual inspection, whereas the red points represent the 198 BCGs used in our final analysis (see section \ref{simulations}).}
\label{fig:lxz}
\end{figure}

\par In total, 329/433 objects passed the visual inspection stage and cuts (see figure \ref{fig:zphotz}). In this analysis, we opted to reject BCGs with obvious tidal features from undergoing a major merger (7/104) as well as ones with obvious image issues, such as exceptionally bright stars, extensive field overcrowding (e.g. for a few fields close to the galactic plane (none of the sample however lie within $-25^{o} \geq \ b \ \leq +25^{o}$) and object truncation due to proximity to field edges (22/104). The remainder were either algorithm `artifacts' (such as a wrongly-attributed CODEX X-ray source to a red sequence detected by \textsc{redMaPPer}, or a projected overdensity of objects with similar photo-z values from \textsc{redMaPPer} but highly variable spec-z values; 14/104) or had no spectroscopic redshift available in the legacy SDSS archive within a 3$"$ match (61/104). A number of example BCGs can be seen in figure \ref{fig:bcgexamples}; all of them have early type morphologies.
\section{Analysis}\label{analysis}
\subsection{Cluster Properties}\label{dynamics}
Three independently measured cluster properties are used to characterise the clusters in our sample: halo mass $M_{200}$ (with respect to the critical density), X-ray luminosity $L_{X}$, and cluster richness $\lambda$, the latter two of which we take directly from the SPIDERS catalogues. The richness values we use originate from \textsc{redMaPPer} and represent the summed \textit{probability} of galaxy membership for a given cluster with redshift taken into account, shown by \cite{2012ApJ...746..178R} to be a superior measure of total membership than imposing a basic colour cut criterion. In addition, the $L_{X}$ values represent the total aperture-corrected luminosity across the entire [$0.5-2$] keV band (see \citealt{2007ApJS..172..182F} for the correction method). The corresponding $L_{X}-z_{\mathrm{RM}}$ distribution for our clusters is shown in figure \ref{fig:lxz}. We limit our clusters at the analysis stage to those above $10^{43} \ \mathrm{ergs \ s^{-1}}$. This was done primarily because such objects constitute poor clusters and low mass groups, often with less well-constrained X-ray measurements in the context of RASS data (e.g. \citealt{2001A&A...369..826B}). If included, however, we observe very little change to our results (less than 5\% in most cases).
\par In this work, we rely on velocity dispersion as a proxy for dynamical cluster mass. The velocity dispersion of a cluster is known to be an effective tracer of the overall mass of the cluster halo (e.g. \citealt{2016MNRAS.462.4117C}), however, this technique is prone to certain biases. For example, in the case of colliding clusters, atypically high galaxy peculiar velocities can also arise due to the cluster being out of dynamical relaxation (as an example, the Bullet cluster has a velocity dispersion of $\sim 1500 \ \mathrm{km s^{-1}}$, see \citealt{2002ApJ...567L..27M}). In addition, interlopers along the line of sight can cause significant problems if not accounted for correctly, which is often a non-trivial task \citep[e.g.,][]{2005AJ....130..968M}. In the context of recent studies, \cite{2016MNRAS.461.4182M} found that cluster masses derived from velocity-based caustics were biased high by $\sim$ 20\%. Although their caustic-based method differs from that used here, their work represents a useful indication as to how much of a deviation one should expect using velocity-based mass proxies.
\par We use a similar prescription in this work to that of \cite{2016MNRAS.463.4490C} when computing the velocity dispersions of our clusters. We make no prior assumptions about the membership of objects contained in the SPIDERS catalogues, requiring only that they meet our quality cuts (see section \ref{sampleselection}). We do, however, apply an additional cut before computing our velocity dispersion estimates to remove projected interlopers, in that we require a galaxy to lie within a projected distance of $R \leq 2 \times R_{200}$ from the designated BCG.
\par For a given cluster member, we follow the standard practice of computing its velocity with respect to the cluster rest-frame \citep[e.g.,][]{1996ApJ...462...32C}:
\begin{equation} \label{eq:velocity}
\frac{v_{i}}{c} \ = \ \frac{z_{i} \ - z_{\mathrm{clus}}}{1 \ + \ z_{\mathrm{clus}}} \ ,
\end{equation}
where $v_{i}$ is the recession velocity of a member galaxy, $z_{i}$ is its corresponding redshift value and $z_{\mathrm{clus}}$ is the redshift of the cluster. We applied an iterative clip to our cluster redshift values in velocity space, imposing a $\pm$3000 $\mathrm{km \ s^{-1}}$ threshold about the median velocity of the cliplist, which was recalculated at each step. We then estimated the final cluster redshift using a biweight. 
\par We apply the same cut in velocity space to our object list, requiring again that a given galaxy be within $\pm$3000 km$ \ \mathrm{s^{-1}}$ with respect to the median cluster velocity; galaxies outside this limit were flagged as interlopers and discarded. The final object list underwent an iterative 3$\sigma$ clip, which was allowed to run until convergence. Clipped object lists with $N < 10$ members were rejected, and a mass was not computed for the cluster (e.g. \citealt{1995MNRAS.274.1071C}).
\par Following the methodology outlined in \cite{1990AJ....100...32B}, we use two different measures of velocity dispersion, dependent on the number of remaining cluster members. For $10 < N \leq 15$, we apply the `gapper' method, which is optimised for clusters with a low member count (see \citealt{1990AJ....100...32B}). For a list of member galaxy line of sight velocities in ranked order, one can use the `gaps' between them to achieve a sense of scale for the underlying distribution: 
\begin{equation}
\sigma_{200} = \frac{\sqrt{\pi}}{n(n - 1)}\sum_{i = 1}^{n - 1}w_{i}g_{i} \ ,
\end{equation}
where $\mathrm{\sigma_{200}}$ is the velocity dispersion and $n$ is the variable rank. The `gaps', $g_{i}$, are defined as: 
\begin{equation}
g_{i} = v_{i + 1} - v_{i} \ ,
\end{equation}
for $1 \leq i < n - 1$, where $v_{i}$ is the velocity of a galaxy ranked at $i$. These gaps are then `weighted' with rank-dependent weights $w_{i}$:
\begin{equation}
w_{i} = i(n - i) \ .
\end{equation}
\par For clusters with a larger number of remaining members ($N > 15$), we adopt a biweight technique as it generally represents a more robust statistic. In both cases, errors on our velocity dispersions are estimated from 68\% confidence limits taken from 10,000 bootstrap realisations, with the mean error $\Delta{{\sigma}_{200}} \sim$ 20\%.
\par For mass estimates, this work follows \cite{2005ApJ...630..206F}, who adopt the following equation for cluster mass:
\begin{equation} \label{eq:m200}
M_{\mathrm{200}} = 1.2 \times 10^{15} \left(\frac{{\sigma}_{200}}{1000 \ \mathrm{km \ s^{-1}}}\right)^{3} \ \\
\frac{1}{\sqrt{\Omega_{\Lambda} \ + \ \Omega_{M}(1 \ + \ \textit{z})^{3}}} \ \mathrm{h_{100}^{-1}M_{\odot}} \ .
\end{equation}
\par
\begin{figure}
\centering 
\includegraphics[width=8.5cm,height=8.5cm,keepaspectratio,trim={2.5cm 0.5cm 3cm 0cm},clip]{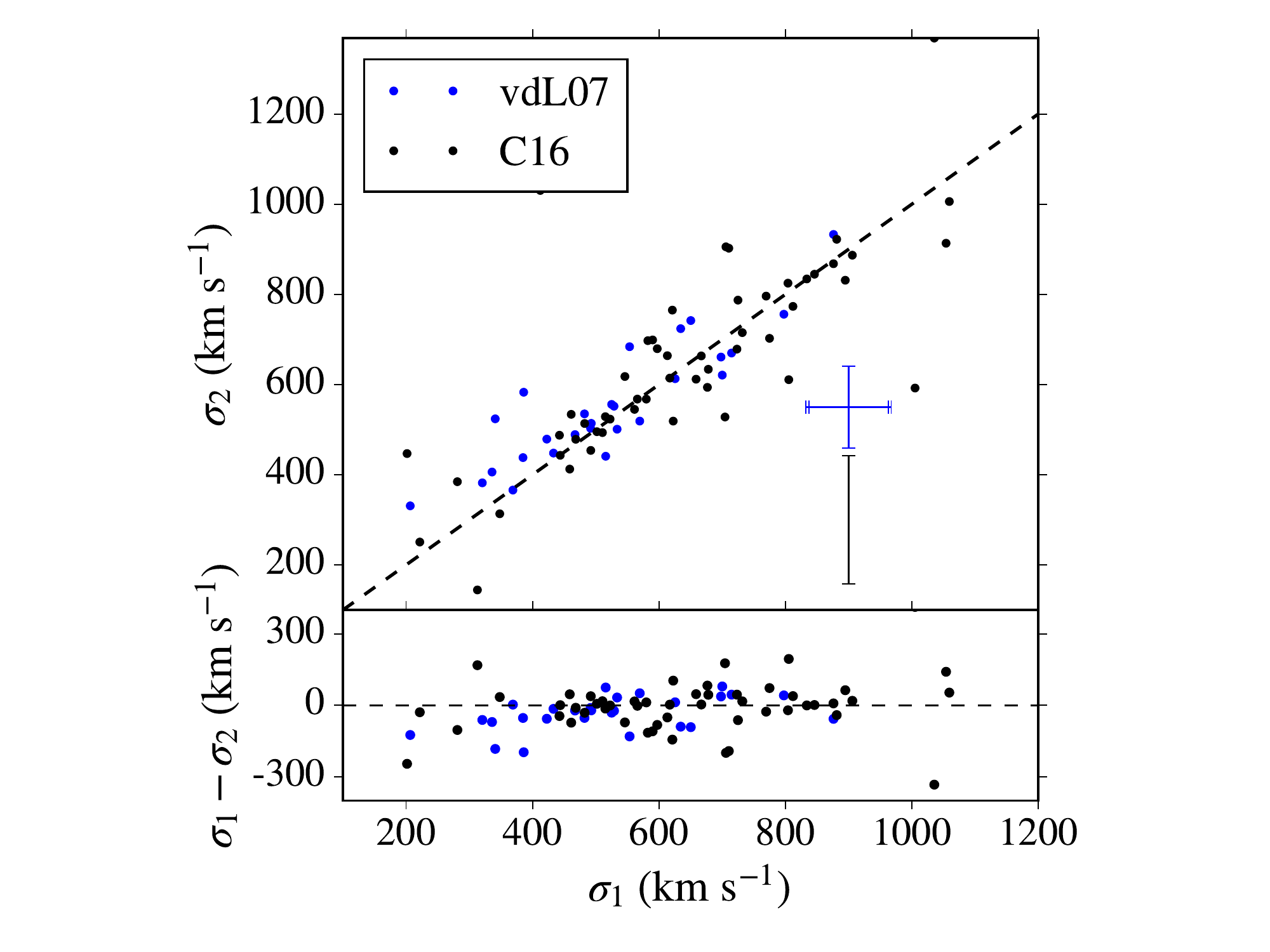} 
\caption{Comparison plot between 81 velocity dispersion values computed here (y-axis) and those computed by \protect\cite{2007MNRAS.379..867V} and by \protect\cite{2016MNRAS.463.4490C} for 27 and 54 common clusters respectively, accounting for duplicates (x-axis). The average $1\sigma$ error for each of the two samples are plotted in the bottom right hand corner.}
\label{fig:sigmacomp}
\end{figure}

 Of the clusters used in this work, 318 have measured velocity dispersions and thereby corresponding dynamical mass estimates. The clusters themselves span a large range of masses, from $10^{13} \ \leq M_{200} \leq \ 10^{15} (\mathrm{M_{\odot}})$ (median $ = 1.4 \times 10^{14} \ \mathrm{M_{\odot}}$; see section \ref{results}). It was found at this stage that a further two BCGs had large deviations in redshift from their corresponding $z_{\mathrm{clus}}$ (SPIDERS ID 1\_6003 and 1\_21735); these clusters were flagged and did not factor into any further analysis. For reference, we present a comparison between ${\sigma}_{200}$ estimates for an overlapping sample of 27 and 54 SPIDERS clusters with \cite{2007MNRAS.379..867V} and \cite{2016MNRAS.463.4490C} respectively in figure \ref{fig:sigmacomp}. In an upcoming paper to be released by the SPIDERS collaboration, we attempt to address these issues through visual screening efforts with the aim of retrieving more accurate estimates of velocity dispersion (Clerc et al. in prep; see \citealt{2016MNRAS.463.4490C} for method).  
\subsection{BCG Structural Parameters: The S\'{e}rsic Profile} \label{sigmadescript}
In order to gain information about the morphological properties of our BCGs, we model their light profiles in two dimensions (see \citealt{2011MNRAS.414..445S} for an example in the context of BCGs). The image data used in this work originate from the SDSS DR12 release (\citealt{2015ApJS..219...12A}). The data have been scaled, preprocessed and have undergone a global background subtraction following \cite{2011AJ....142...31B}. The background treatment is much improved over the sigma-clipping method used in releases prior to DR8 (see fig. 12 from \citealt{2011AJ....142...31B} for comparison), which is known to over-subtract the profile wings of high-S\'{e}rsic index objects such as BCGs (e.g. \citealt{2007AJ....133.1741B}). Here, we choose to use these estimates for the global sky during fitting, which we discuss in section \ref{sigpipedescript}.
\par The BCGs in this work are modelled by a single-S\'{e}rsic profile (\citealt{1963BAAA....6...41S}), which has the form  
\begin{equation} \label{eq:1}
I(R) \ = \ I_{e}\exp{ \{ b_{n}[(R/R_{e})^{1/n} \ - \ 1] \}} \ ,
\end{equation}
where $I(R)$ is the surface brightness of an object at radius $R$, $R_{e}$ is the effective radius within which 50\% of the object light is contained, $I_{e}$ is the object surface brightness at the effective radius, $n$ is the S\'{e}rsic index and $b_{n}$ is a product of incomplete gamma functions as described in \cite{1991A&A...249...99C}. The empirical S\'{e}rsic profile is a good description of the light distribution of BCGs, which are predominantly bulge-dominated and ellipsoidal in nature (see \citealt{2005PASA...22..118G} and references therein). The integrated magnitudes produced from S\'{e}rsic fitting have the additional advantage that they account for more light than traditional Kron/Petrosian magnitudes for objects with high-$n$, the latter of which can underestimate the true brightness by as much as 95\% (\citealt{2005PASA...22..118G}).
\par For the purposes of this study, we opt to fit a single-S\'{e}rsic component to each BCG. There is some debate in the literature as to the most appropriate number of Sersic components which should be fit to a galaxy in order to accurately estimate galaxy parameters. For example, \cite{2014MNRAS.443..874B} experimented with fitting single- and double-S\'{e}rsic fits to a series of simulated single- and double-S\'{e}rsic galaxy images. They found that both the single- and double-S\'{e}rsic fits perform well at the bright end ($M_{r} > -21$) regardless of galaxy type, with both models increasingly ill-suited to their opposing galaxy type at fainter magnitudes. It was shown that a single-S\'{e}rsic fit to what is, in reality, a double-S\'{e}rsic galaxy is as inappropriate a model as applying a double-S\'{e}rsic fit to what is, in reality, a single-S\'{e}rsic galaxy. On closer inspection of the light profiles for most of our BCGs, we found no significant evidence for the addition of a secondary component, so therefore adopt a single-S\'{e}rsic as our model type to fit to each BCG.
\par Numerous profile-fitting codes exist in the literature, such as: \textsc{GIM2D} (\citealt{1998ASPC..145..108S}), \textsc{ProFit} (\citealt{2017MNRAS.466.1513R}), \textsc{Imfit} (\citealt{2015ApJ...799..226E}), \textsc{PyMorph} (\citealt{2010MNRAS.409.1379V}) and \textsc{GALFIT} (\textsc{GALFIT3}, \citealt{2010AJ....139.2097P}). For this work, we opt to use \textsc{GALFIT}, which uses a Levenberg-Marquardt (`downhill-rolling') based ${\chi}^{2}$ minimisation technique for fitting light profiles (see \citealt{2016MNRAS.460.3458K} for a recent use of the software). It is also capable of generating light profiles based on fixed input parameters, which we use to create mock galaxies as described in section \ref{simulations}. 
\par We run \textsc{GALFIT} using the Structural Investigation of Galaxies via Model Analysis code, or \textsc{SIGMA} (\citealt{2012MNRAS.421.1007K}), which is an R-based pipeline software used with great success in the GAMA survey (\citealt{2011MNRAS.413..971D}) to provide S\'{e}rsic model fits to 167,600 galaxies in five optical (SDSS$-ugriz$, \citealt{1996AJ....111.1748F}) and 4 near-infrared (UKIRT$-YJHK$) passbands (see \citealt{2011MNRAS.412..765H}). \textsc{SIGMA} is capable of performing a full fit, including: object extraction through \textsc{Souce Extractor} (\textsc{SExtractor}; \citealt{1996A&AS..117..393B}), creating a model of the field PSF, estimating the local sky about an object, masking external objects and, finally, fitting a profile through \textsc{GALFIT}. Details of our implementation of \textsc{SIGMA} are provided in the upcoming section; more specific information can be found in \cite{2012MNRAS.421.1007K}. 
\subsection{\textsc{SIGMA} Pipeline and Implementation}\label{sigpipedescript}
\par \textsc{SIGMA} is designed to fit objects in multiple bands in tandem, requiring only the coordinates at which an object is located as an input file alongside any image data corresponding to the desired fitting bands. \textsc{SIGMA} produces an output file containing any relevant \textsc{SExtractor} parameters used in fitting, the background estimate (if requested), and the profile-fitting results from \textsc{GALFIT}. A description of the pipeline is given below, alongside any parameter settings used here when running \textsc{SIGMA}.
\par \textit{i. Image Cutout}: To begin, \textsc{SIGMA} accesses the WCS information in the header of an image file containing an object. It then converts the celestial RA/DEC coordinates of an object into $x/y$ cartesian pixel coordinates using the \textsc{sky2xy} routine in the \textsc{WCSTools} package (\citealt{1998DDA....29.0806M}). The upper and lower limits of a 1201$\times$1201 pixel region centered on the primary object are determined. This cutout, designated the `primary science image' is used during analysis from this point forward.
\par \textit{ii. Source Extraction}: \textsc{SIGMA} then runs \textsc{SExtractor} on the primary science image to detect any objects in the field. During extraction we applied the default \textsc{SIGMA} parameters: \textsc{DETECT\_THRESH} = $2\sigma$, \textsc{DETECT\_MINAREA} = 10 and \textsc{SATUR\_LEVEL} = 25,000. The image is also filtered through a 5$\times$5 pixel Gaussian convolution kernel with FWHM = 2 pixels prior to detection; \textsc{SExtractor} defaults are used everywhere else. Outputs from \textsc{SExtractor} corresponding to the primary object provide initial parameter estimates for the \textsc{GALFIT} algorithm. Any extra object detections about the primary object are designated `secondary' sources. The number of secondaries to be used during fitting is specified by the user prior to running \textsc{SIGMA}; any other objects are designated as `tertiaries' and masked out prior to fitting (see \textit{vi}.). 
\par \textit{iii. PSF Extraction}: The output catalogue from \textsc{SExtractor}is used to find objects in order to estimate the PSF of the field. The \textsc{PSF Extractor} software package (\textsc{PSFEx}; \citealt{2013ascl.soft01001B}) is applied to create the empirical PSFs used in \textsc{SIGMA}, which are smoothed by fitting the end result with a Moffat profile (\citealt{1969A&A.....3..455M}). For an object to qualify for  computing the PSF, \textsc{SIGMA} requires a S/N > 10 and an eccentricity, $\epsilon$, of > 0.05. \textsc{PSFex} then estimates the FWHM of the object, requiring that 2 < FWHM < 10 pixels and that the object lies within the central 50\% of the distribution. A minimum of 10 objects are required to compute a PSF; if this criterion is not met, \textsc{SIGMA} will loop back and attempt to run \textsc{PSFEx} on a larger cutout area. Upon completion, the final result is a 25 $\times$ 25 pixel PSF estimate, which is used in fitting with \textsc{GALFIT}.
\par \textit{iv. Sky Subtraction}: To estimate the local sky about an object, \textsc{SIGMA} uses an adaptive mesh technique dependent on the size of the primary object (see section 3.3 of \citealt{2012MNRAS.421.1007K}). Within each `cell' of the mesh, \textsc{SIGMA} computes the median background level and fits a smooth spline across the frame, subtracting the background estimate from the science image. Although suitable for objects in the field/groups, we prefer not to apply any additional background subtraction to our images due to the nature of the cluster environment. In the cores of clusters, it is often the case that galaxies have overlapping profile wings. In addition, cluster cores also have a faint ICL component, thought to have formed primarily from stars ejected from galaxies through mergers, tidal stripping, and harassment (\citealt{2007ApJ...666..147G}; \citealt{2005ApJ...631L..41M}; \citealt{2012MNRAS.425.2058B}). Therefore, we opt to use the global SDSS estimates when fitting our objects to avoid any sky bias at local scales, primarily as it is known that sky overestimation has a significant impact on the final fit of an object (e.g. \citealt{2005PASA...22..118G}). We explore the impact of field-by-field variations on object fitting in our simulations (section \ref{simulations}).
\begin{figure}
\centering 
\includegraphics[width=8.7cm,height=8.7cm,keepaspectratio,trim={2.5cm 0cm 2cm 0cm},clip]{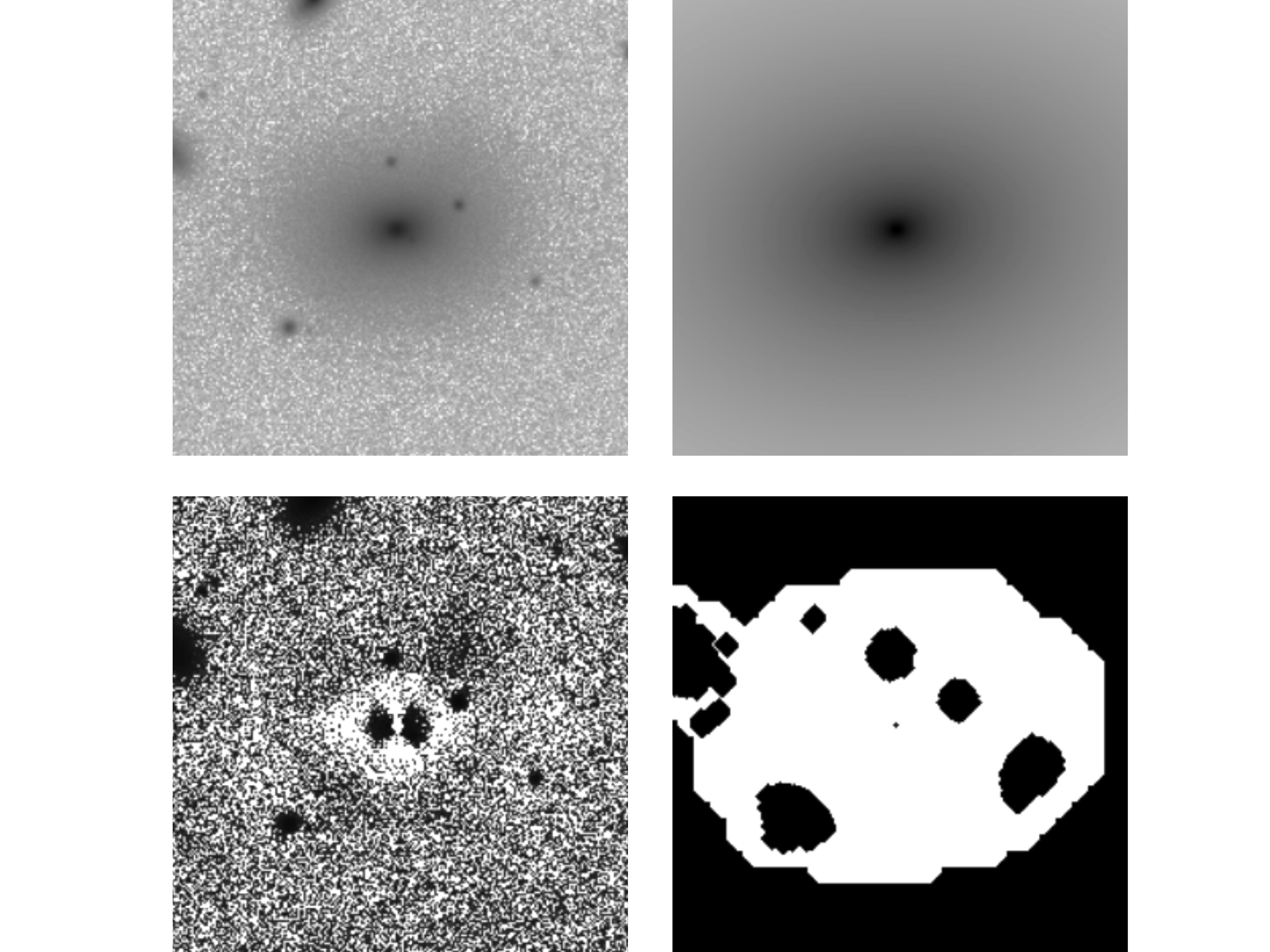} 
\caption{An example data product from \textsc{SIGMA} for the BCG in cluster 1\_1365 (100 $\times$ 100 kpc, $z$ $\sim$ 0.05). Clockwise from top left: SDSS $r$-band image, S\'{e}rsic model from \textsc{GALFIT}, image mask, residual (enhanced to bring out faint artifacts). Except for the binary mask, all other quadrants are scaled logarithmically.}
\label{fig:exampledataprod}
\end{figure}
\par \textit{v. Object Masking}: \textsc{SIGMA} produces segmentation maps that can be used as object masks; the mask shape and the number of sources to be used during fitting are specified by the user. In this work, we apply inner (to minimise the effect of PSF-sensitive cores on our fits), outer, and tertiary masking to our objects, using a pixel buffer of 5 and a 1 pixel mask to the inner region. We impose a maximum limit of three unmasked secondary sources within 2.5 magnitudes of the primary source; we choose this value primarily as a compromise between computational efficiency and resolving parameter degeneracies at the fitting stage. 
\par \textit{vi. Profile Fitting}:
Prior to fitting, \textsc{SIGMA} creates a further cutout of the science image based on output parameters from \textsc{SExtractor} for the primary object. As aforementioned, initial guesses for both the primary object and secondary objects are taken from \textsc{SExtractor}, including: object magnitude, axis ratio, position angle and effective radius $R_{e}$ (see \citealt{2012MNRAS.421.1007K} for details of the effective radius estimation). The initial estimate of the S\'{e}rsic index $n$ is set to a constant value of 2.0; however, changing this value was found by \cite{2012MNRAS.421.1007K} to have little effect on the final result. We modelled our objects with a free S\'{e}rsic profile (1 < $n$ < 20), imposing no manual constraints beforehand on any output parameters (see discussion below). 
\par In order to impose quality control on the \textsc{GALFIT} output values, \textsc{SIGMA} contains inbuilt constraints which are then applied post-fitting. After running \textsc{GALFIT} on an object, \textsc{SIGMA} analyses the output and refits any objects which are flagged as follows: 
\begin{enumerate}
  \item \textsc{GALFIT} has encountered a serious error that prohibits completion of the fit.
  \item The primary object's centre has undergone a migration of $x^{2}+y^{2}$ > $R_{e, \mathrm{initial}}$.
  \item The primary object's end radius is either exceptionally large ($\mathrm{log_{10}}\left({\frac{R_{e, \mathrm{final}}}{R_{e, \mathrm{initial}}}}\right) > 3 $) or exceptionally small ($\mathrm{log_{10}}\left({\frac{R_{e, \mathrm{final}}}{R_{e, \mathrm{initial}}}}\right) < -3 $).
  \item The primary object's end ellipticity is high ($\epsilon > 0.95$).
\end{enumerate}
In total, 326/327 objects were fit with profiles from \textsc{SIGMA}; we discuss the use of the $i$-band model profiles when computing stellar masses in section \ref{stellarmass} and our choice of the $r$-band for the morphological parameters. Of these, 198 galaxies were used in our subsequent analysis, which we discuss in the upcoming section. An example of the finished \textsc{SIGMA} data product for a BCG is shown in figure \ref{fig:exampledataprod}.
\subsection{Probing Surface Brightness Limits with Simulated Profiles}\label{simulations}
As our BCGs encompass a large magnitude range, we decided to investigate any potential structural dependencies that may arise with changes in surface brightness or $R_{e}$/$n$. Little investigation has been done to understand such biases which arise a result upon fitting (see \citealt{2009MNRAS.398.1129G}, \citealt{2011AJ....142...31B}, \citealt{2017arXiv170602704V} as examples of studies that attempt to characterise structural biases), with numerous authors content to base their fit quality on the reduced ${\chi}^{2}$ estimate (${{\chi}_{\nu}}^{2}$). These fits are not necessarily representative of a result which is physically meaningful (e.g., \citealt{2008MNRAS.385...23L} found differing results for their BCGs, dependent on the isophotal level chosen or method of background subtraction). Some studies have even adopted a novel approach to their fitting in order to gain insight into the `goodness of fit' of their objects; a recent example being \cite{2016MNRAS.462.1470L}, who fit their objects with \textsc{SIGMA} using a Monte-Carlo based method, taking the resulting parameters for each fit at the point of convergence. 
\par We therefore tested \textsc{SIGMA}'s ability to recover parameters from BCG-like objects with the fitting criteria specified in section \ref{sigmadescript} by creating a grid of model profiles with known input parameters, which we then inject into real SDSS fields. The 8 SDSS fields selected for use in the simulations are listed in table \ref{t:fields}. Although they were taken from among the fields in the sample, the selection process was effectively at random. The fields contain many features known for causing issues during fitting, such as bright stars and regions of high source density.
\begin{table}
\centering
\begin{tabular}{rccccc}
\noalign{\smallskip} \noalign{\smallskip
}
\hfill ID & \hfill ${\alpha}_{2000}$ & \hfill ${\delta}_{2000}$ & \hfill RUN & \hfill CAMCOL & \hfill FIELD	\\
\hline
1\_1282        & 133.761    & +55.454  &	1350	& \hfill	2	&	\hfill 194	\\
1\_14572       & 352.634	& +20.728  &	8096	& \hfill	6	&	\hfill 182	\\
1\_2785        & 143.215	& +47.931  &	2740	& \hfill	6	&	\hfill 261	\\
1\_2952        & 163.487	& +49.499  &	2883	& \hfill	1	&	\hfill 109	\\
1\_3349        & 202.598	& +49.185  &	3650	& \hfill	5	&	\hfill 85	\\
1\_4285        & 146.506	& +43.127  &	2887	& \hfill	4	&	\hfill 251	\\
2\_11151       & 11.534		& +20.614  &	7913	& \hfill	5	&	\hfill 30	\\
2\_2401        & 126.401	& +48.341  &	1331	& \hfill	4	&	\hfill 156	\\
\noalign{\smallskip} \noalign{\smallskip}
\end{tabular}
\centering
\caption{Cluster fields used when inserting mock galaxies. The ID column refers to the corresponding ID string in the SPIDERS catalogues.}
\label{t:fields}
\end{table}
\begin{figure}
\centering 
\includegraphics[width=8.7cm,height=8.7cm,keepaspectratio,trim={3cm 1cm 2cm 0cm},clip]{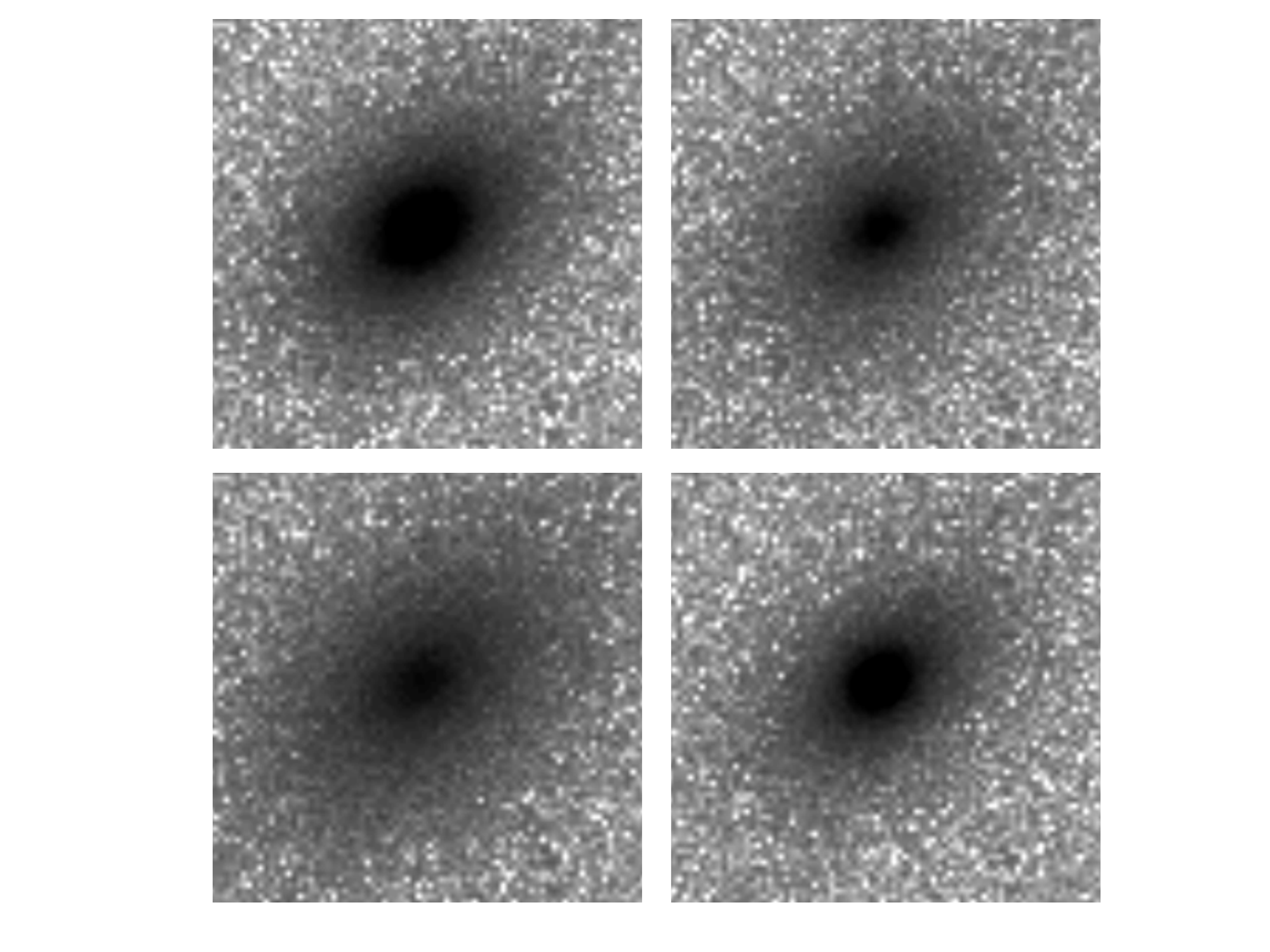} 
\caption{Example $r$-band simulated profiles at $15^{\mathrm{th}}$ magnitude (tiles scaled to 100 $\times$ 100 kpc with an assumed redshift of 0.2). Top: for a fixed $n$ of 4, $R_{e}$ = 20, 80 kpc. Bottom: for a fixed $R_{e}$ of 40 kpc, $n$ = 2, 8. Note the change in surface brightness distribution in either case.}
\label{fig:example profiles}
\end{figure}
\par The model profiles were generated using \textsc{GALFIT} for consistency, as it was the algorithm of choice used during fitting. The estimated magnitude zeropoints and PSFs (from \textsc{SIGMA}) for these fields were used when generating the model profiles, to account for field-by-field photometric variations. An idealised Poisson noise component was also added, to mimic shot noise. The models were then injected into each of the 8 fields at 8 random, fixed positions and subjected to a full run through the \textsc{SIGMA} pipeline. We chose a range of S\'{e}rsic indices from $1-10$ in steps of 1, apparent magnitude values from $12-19$ in all three bands and effective radii from $10-100$ kpc at $z = 0.2$ (equivalent to 0.253 $"$/kpc). Therefore, a total of 6,400 model combinations per band were created, or 19,200 in total. We chose this scale for $R_{e}$ as it represents the approximate peak of the SPIDERS sample, so will correspond to effective radii relevant for our purposes. In addition, we justify our selection of input S\'{e}rsic indices by noting that they encompass the bulk of all potentially physically-meaningful outputs; S\'{e}rsic indices larger than $n = 10$ represent virtually identical profiles (e.g. \citealt{2005PASA...22..118G}). Finally, the input axis ratio was held constant at 0.66 (a realistic value for most BCGs; most SPIDERS BCGs have axis ratios between 0.6-0.8 with little variation across bands) and the position angle at $50^{\circ}$ (not significant in the context of this study); this was primarily done for the sake of preserving computational resources, so to reduce the number of potential parameter combinations to simulate. 
\begin{figure*}
\centering
\includegraphics[width=21cm,height=21cm,keepaspectratio,keepaspectratio,trim={1.5cm 0cm 2.5cm 0cm},clip]{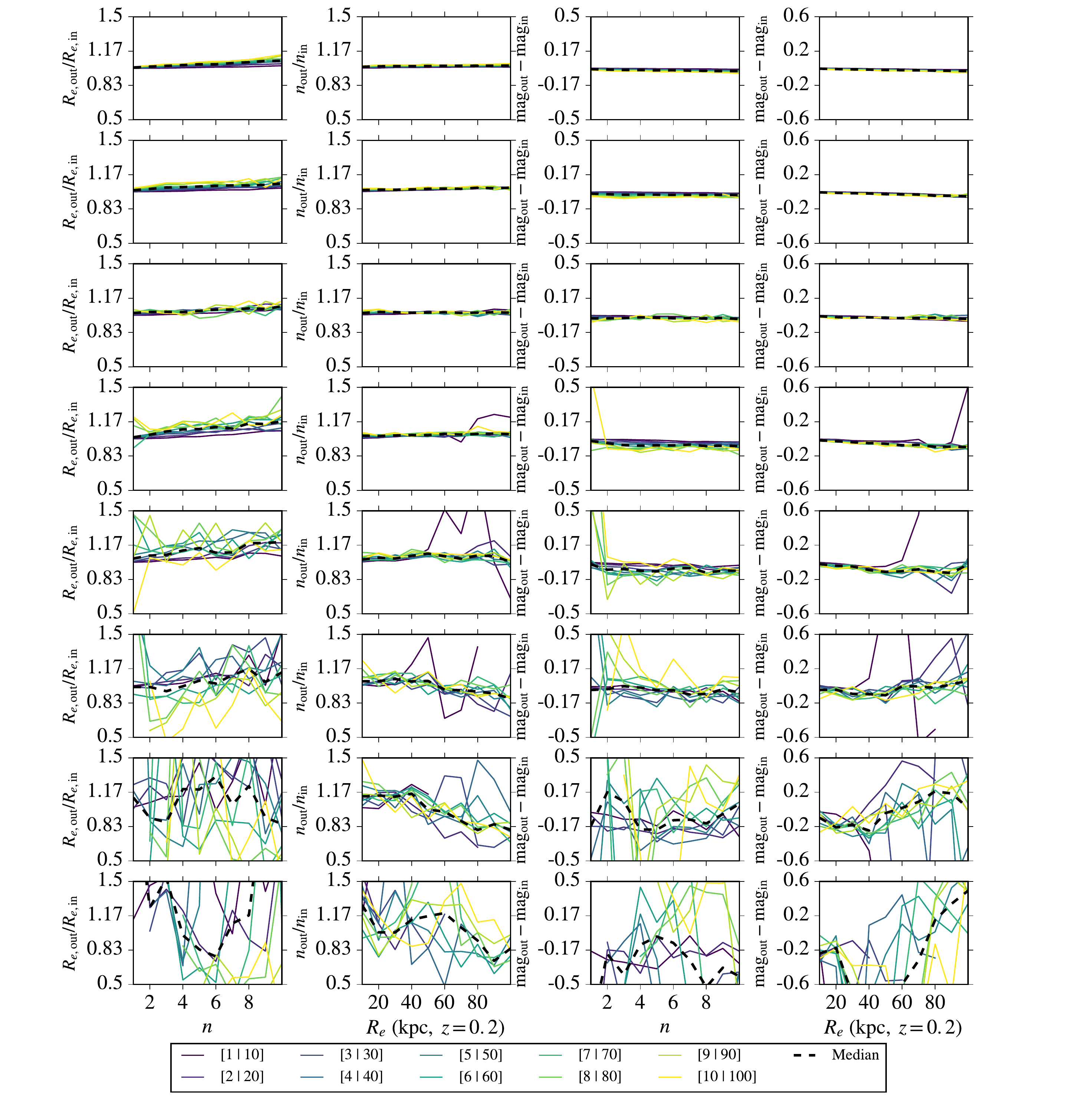}
\caption{The results of fitting simulated profiles with \textsc{SIGMA} in the $r$-band. Each of the four horizontal panels represents the output for one of the eight fixed input magnitudes ($12-19$, descending). The coloured lines represent the biweighted average across the runs (assuming at least 3/8 models at each point were successfully fit; corresponding legend key [$n$ | $R_{e}$] for a given colour, interchange for the relevant panel). The black dashed line represents the median [$n$ | $R_{e}$] of the coloured lines. There is a clear decline in output-to-input S\'{e}rsic index with effective radius at $17^{\mathrm{th}}-19^{\mathrm{th}}$ magnitude (second column).}
\label{fig:simr}
\end{figure*}
\par We refer the reader to the Appendix for the simulation results in the $g$ and $i$ bands, as we choose to use the $r$-band profile parameters for our BCGs. The full outputs for the $r$-band run are shown in figure \ref{fig:simr}, for which we note several initial observations. Firstly, that there is an obvious co-dependency between $n$ and $R_{e}$; the outputs are affected if either is changed (e.g., \citealt{1996ApJ...465..534G}). Brighter than $17^{\mathrm{th}}$ magnitude, the effect is minimal, with an average scatter about the input of $ \sim 5 \%$ in S\'{e}rsic index for a given stepwise change in effective radius and $\sim 20 \%$ vice-versa in effective radius for each incremental change in S\'{e}rsic index. In general, magnitude values are accurately recovered; we see an average scatter of $\sim$ 0.1 magnitudes about the input value, regardless of whether $n$ or $R_{e}$ is changed. As is visible however, the scatter in the values of $n$ and $R_{e}$ becomes more significant at fainter magnitudes (note the bottom two panels of figure \ref{fig:simr}), with the output scatter well in excess of 50\% at $19^{\mathrm{th}}$ magnitude. 
\par The behaviour of the $n = 1$ profile with $R_{e}$ is significantly different than for profiles where $n > 1$. At increasing effective radii, it appears that fits to the profile become `chaotic' at $\sim 14^{\mathrm{th}}$ magnitude, with the scatter exceeding 50\% at $\sim 14^{\mathrm{th}}$ magnitude or fainter. Such a profile is exponential by definition and is often used as a model for disc galaxies (e.g. \citealt{1970ApJ...160..811F}); indeed, it is the default model in the SDSS (e.g., \citealt{2011AJ....142...31B}). The difficulty in fitting such extended, exponential profiles may arise from issues with additional source blending, due to the fact that $n = 1$ profiles are not strongly centrally concentrated (e.g. figure \ref{fig:example profiles} shows the effect of changing the S\'{e}rsic index). However, this profile is generally not appropriate for galaxies such as BCGs, the majority of which are bulge-dominated and follow a de Vaucouleurs-like profile (e.g., \citealt{1948AnAp...11..247D})where $n \sim 4$ (e.g. \citealt{2011MNRAS.414..445S}). Some studies have modelled cD-type BCGs with a bulge+exponential component, in order to account for their extended stellar halo which is degenerate with the intracluster light (e.g. \citealt{2015MNRAS.448.2530Z}, \citealt{2011ApJS..195...15D}).
\begin{figure}
\centering
\includegraphics[width=8.5cm,height=8.5cm,keepaspectratio,trim={0cm 1cm 0cm 0.5cm},clip]{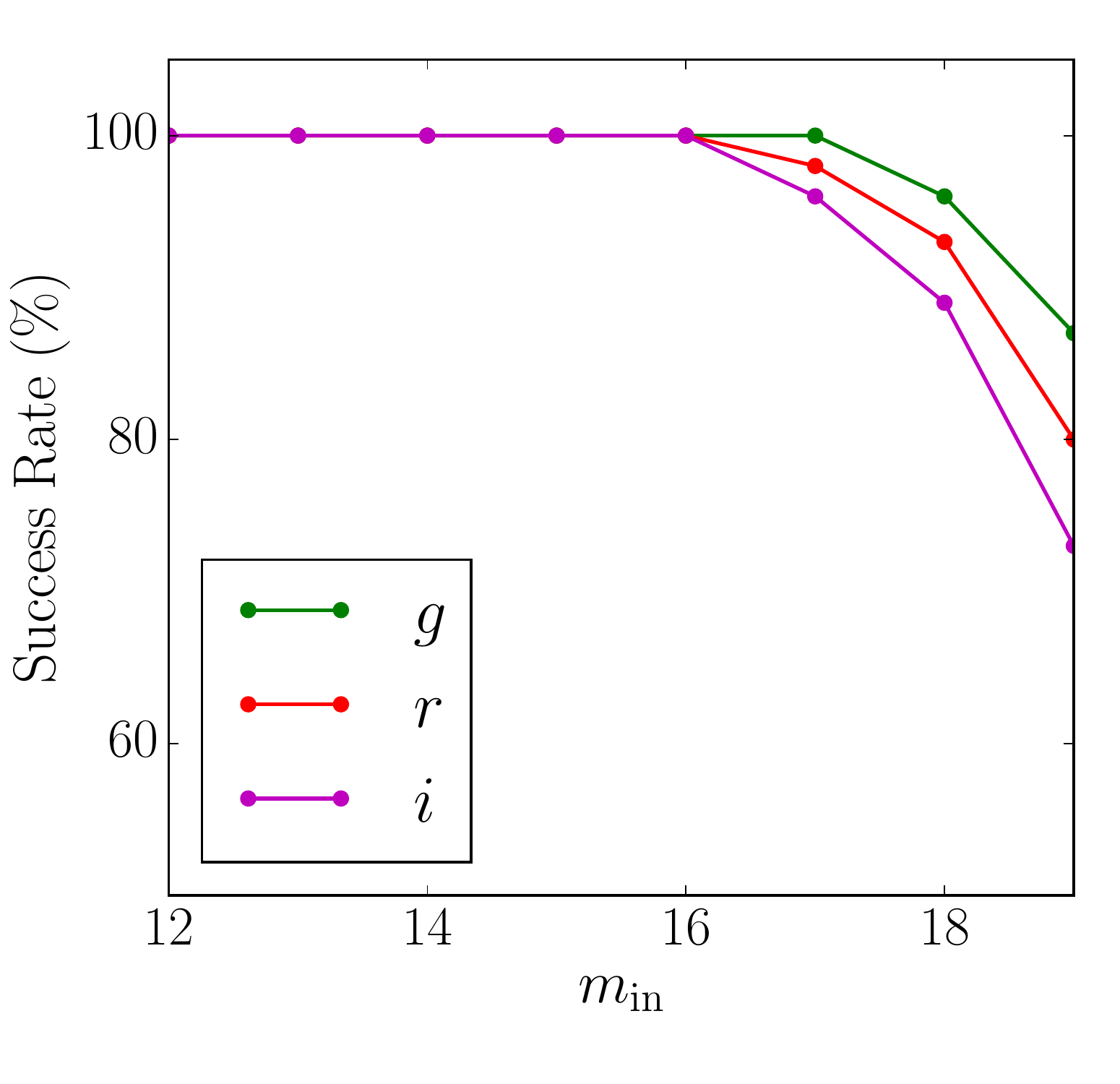}
\caption{The fitting success rates from \textsc{SIGMA} fitting of mock profiles. \textsc{SIGMA}'s performance clearly degrades with both wavelength and input magnitude.}
\label{fig:simcomplete}
\end{figure}
\begin{figure}
\centering 
\includegraphics[width=8.5cm,height=8.5cm,keepaspectratio,trim={1.5cm 0cm 3.5cm 0cm},clip]{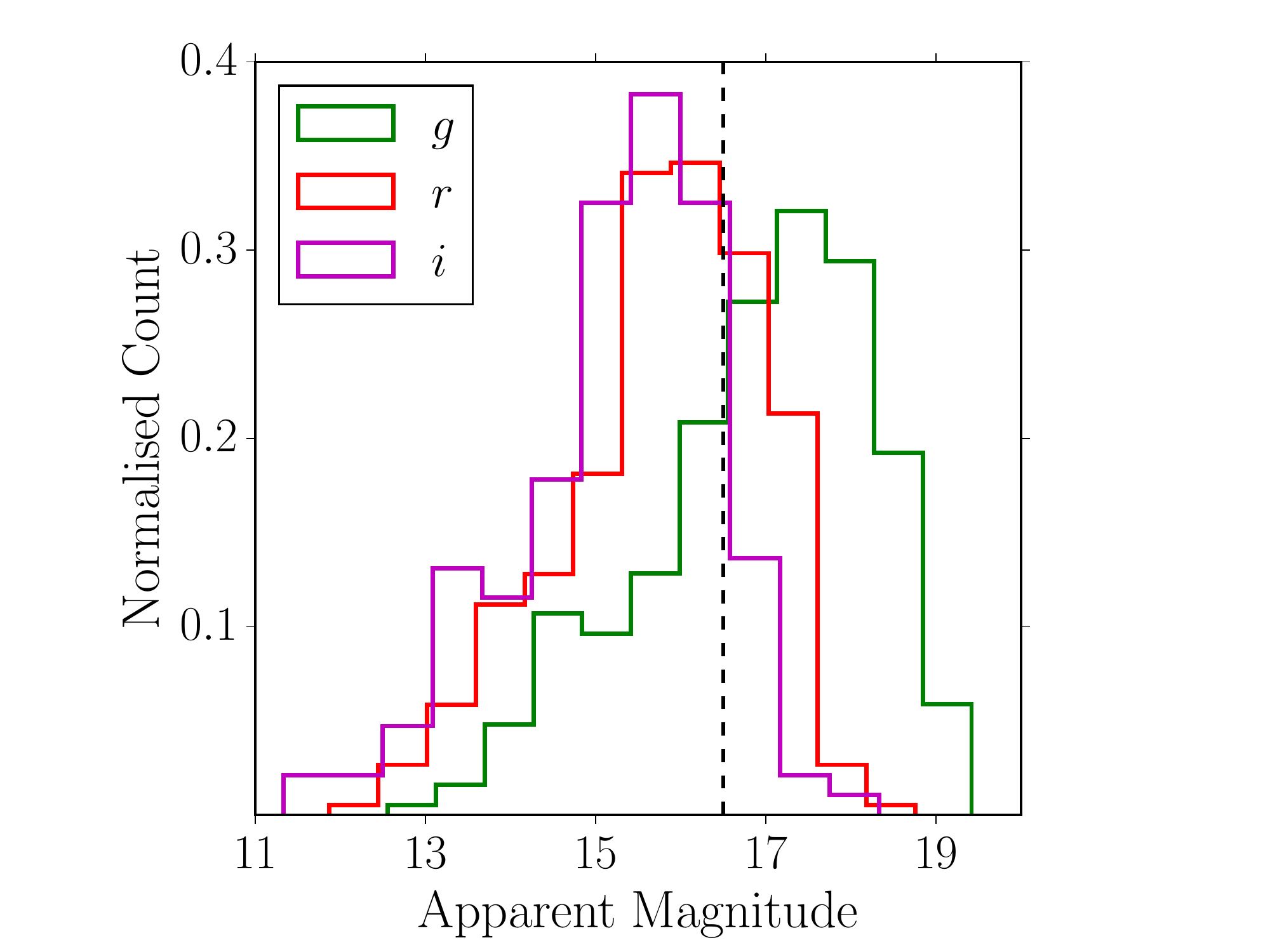} 
\caption{The raw fit magnitudes for the SPIDERS sample from \textsc{SIGMA} (N=326), prior to any k-correction or dust correction. The histograms have been normalised by area.}
\label{fig:rawmags}
\end{figure}
\par From our simulations, we are also able to characterise several important biases, which may have significant consequences if one were to use SDSS data when profile fitting, or when visually-classifying morphologies. There is an obvious bias in S\'{e}rsic index with increasing magnitude. At $\geq 17^{\mathrm{th}}$ magnitude, there is a downturn in output S\'{e}rsic index with increasing effective radius (second panel from the left, figure \ref{fig:simr}), with the reverse being true for effective radius. We strongly suspect this result is due to a surface brightness effect; as the wings of an object become faint with respect to the background level, they become more difficult to detect (surface brightness is dependent on both S\'{e}rsic index and effective radius). Therefore, \textsc{GALFIT} underestimates the true slope of the light profile. A similar effect was also reported in \cite{2013MNRAS.435..623V} for a sample of >3000 galaxies modelled with artificially-redshifted SDSS photometry. This effect is consistent regardless of S\'{e}rsic index, with all of the output values (bar $n$ = 1) showing a similar reduction from input as a trend with effective radius. The effect is also independent of filter; all three bands used for fitting showed a similar bias, albeit to varying degrees of severity (see figures \ref{fig:simg} and \ref{fig:simi} in the Appendix, plus discussion below). 
\par Figures \ref{fig:simg} \& \ref{fig:simi}, also reveal a marked decrease in fit robustness moving from the $g$ to the $i$-band, with the scatter in output values increasing with wavelength. This behaviour is also seen in the decreasing success rates of our simulations (with `success' defined as a given combination of fit parameters being modelled by \textsc{SIGMA} to completion), as presented in figure \ref{fig:simcomplete}. The sky tends to be brighter at redder wavelengths (see \citealt{2002AJ....123..485S}), which is reflected in the smaller magnitude limit of the SDSS $i$ band (21.3 magnitudes, as opposed to 22.2 for $g$/$r$). The increased sky brightness in the $i$ band therefore significantly affects the final fit. We are therefore justified in our selection of the $r$-band models to characterise our BCGs, as a compromise between both band depth and the amount of k-correction necessary (section \ref{stellarmass}).
\par For these reasons, we decided to impose a cut on the raw output magnitudes for the SPIDERS BCGs, to minimise potential biases in S\'{e}rsic index that may arise as a result of surface brightness dimming effects. As we are unsure at the resolution of our simulations precisely where this effect begins to dominate our fits, we select a magnitude of $m_{r} = 16.5$ as the faintest magnitude for which to include objects. To place this choice into context with the output magnitudes from \textsc{SIGMA} of the BCGs in our sample, the raw output magnitude distributions are displayed in figure \ref{fig:rawmags}. The sample peaks at $m_{r} = 15.8$ (with $m_{g} = 16.9$ and $m_{i} = 15.2$), with $\sim 1/3$ of all galaxies lying above the our limiting magnitude of 16.5 (99/326). Thus, the total number of objects for which we include in our an analysis to 227 (198 after imposing further quality cuts as described in section \ref{results}).
\par Applying this cut in BCG magnitude at $m_{r} = 16.5$ significantly affects the redshift distribution of our clusters (figure \ref{fig:lxz}), in that it reduces the number of intermediate-redshift clusters in the sample (notably, between $0.2 < z < 0.3$; this is also revealed by a K-S test ($\mathrm{log_{10}}[p_{\mathrm{KS}}] = -4.56$). However, we detect no evidence for any differences in the distributions of $L_{\mathrm{X}}$, $M_{200}$ and $\lambda$ between the full and magnitude-restricted samples beyond the $1.3 \sigma$ level ($\mathrm{log_{10}}[p_{\mathrm{KS}}] = $ $-0.772$, $-0.491$ and $-0.077$ respectively).
\section{Results}\label{results}
\subsection{Stellar Masses and Structural Parameters}\label{stellarmass}
\begin{figure}
\centering 
\includegraphics[width=11.5cm,height=11.5cm,keepaspectratio,trim={4cm 0.5cm 4cm 0.5cm},clip]{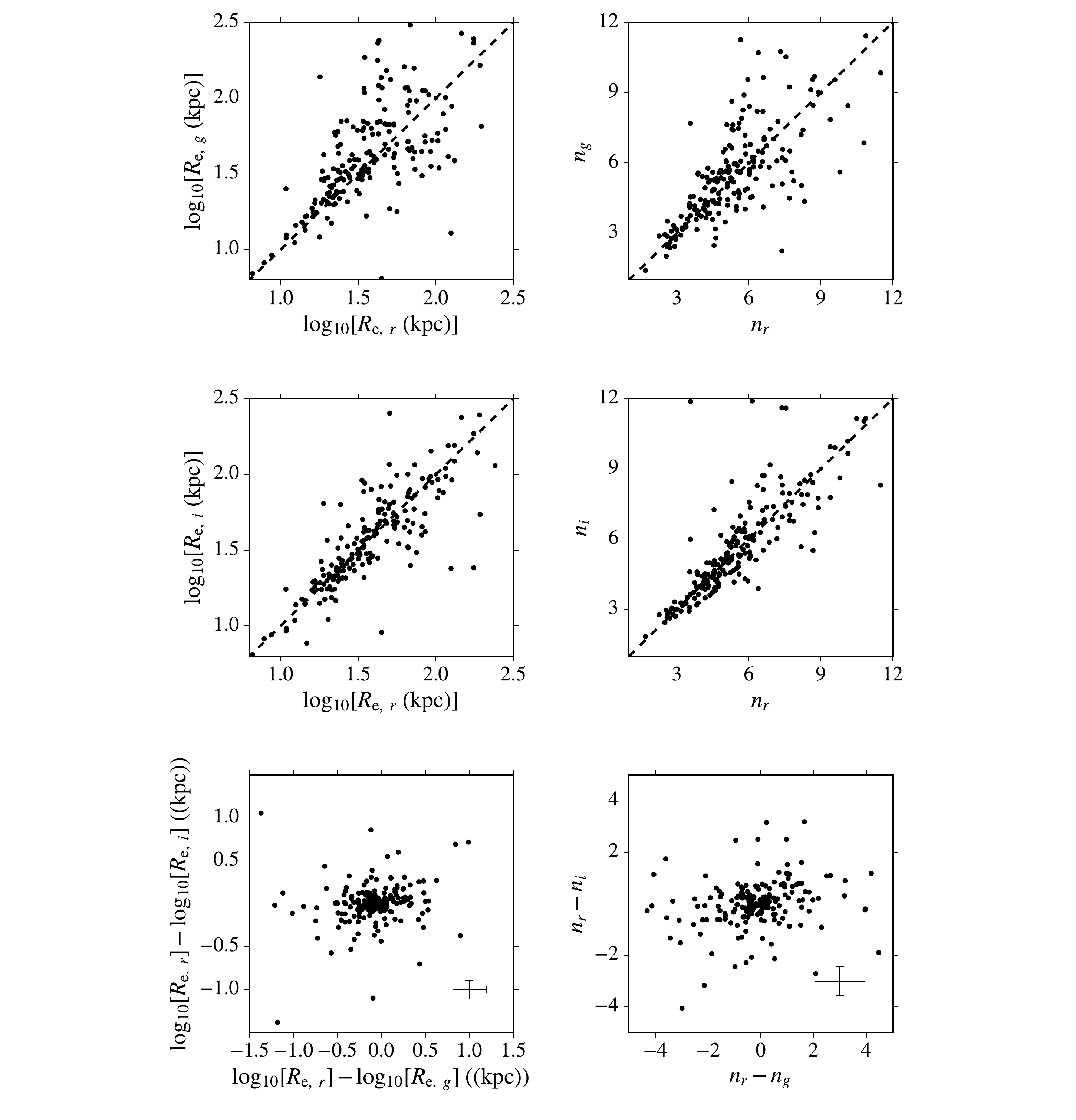} 
\caption{Comparison between the output structural parameters for the BCG sample (N=198). The dashed line represents the 1:1 relation and the crosshairs in the bottom two plots represent the $1\sigma$ scatter in $x$/$y$. In general, there is a good agreement across bands, albeit with increased scatter at large $n/R_{e}$. This has also been demonstrated in our simulations (section \ref{simulations}).}
\label{fig:re_n_comp}
\end{figure}
\begin{figure}
\centering 
\includegraphics[width=8.5cm,height=8.5cm,keepaspectratio,trim={2cm 0cm 2.5cm 0cm},clip]{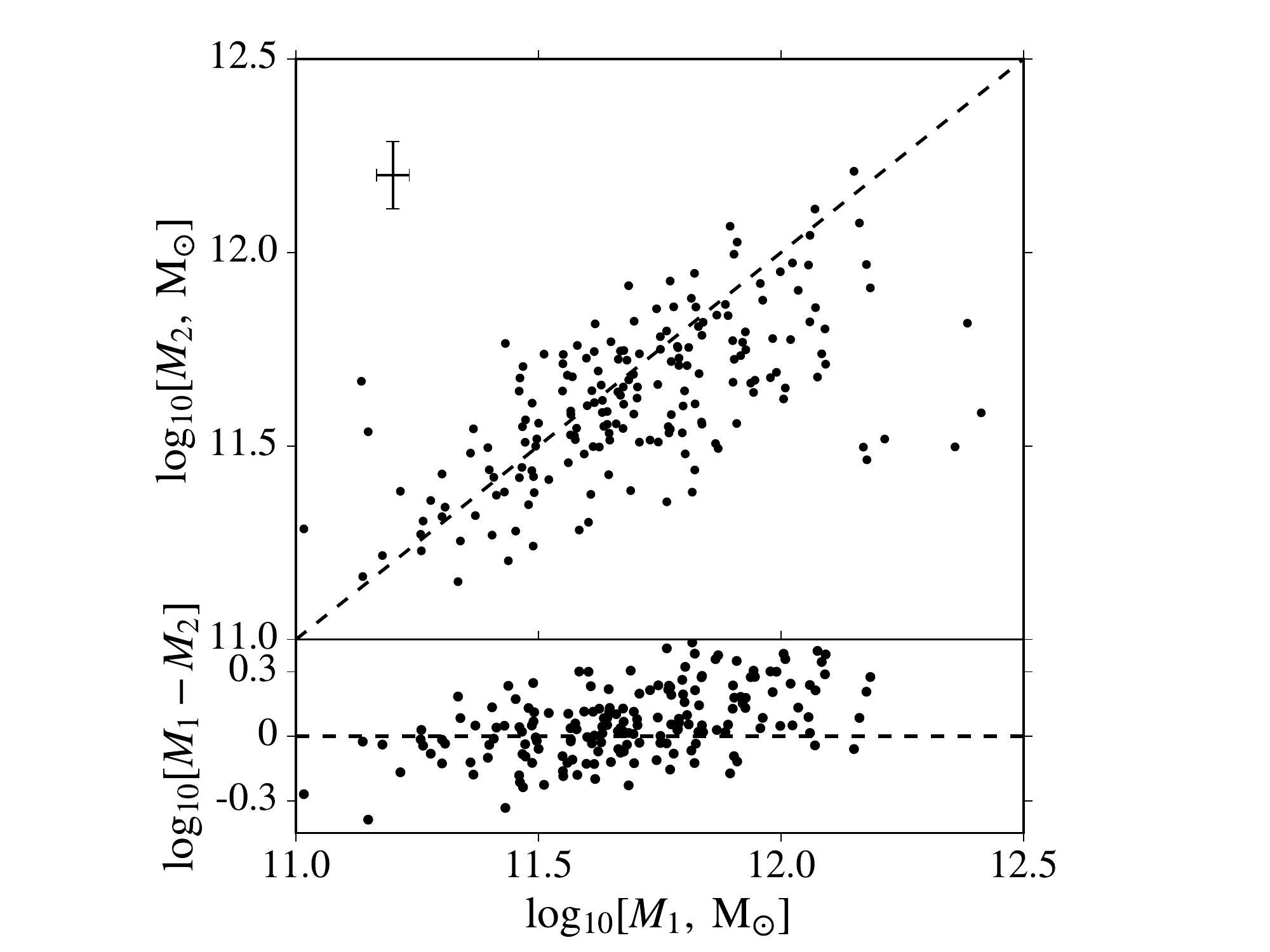} 
\caption{Comparison between the stellar mass estimates for our BCGs using a scaling relation from \protect\cite{2011MNRAS.418.1587T} and those in the MPA-JHU value-added catalogue, for a common sample of 192 BCGs ($x$-axis and $y$-axis, respectively). The crosshairs give the typical error value in each respective axis (for the MPA-JHU catalogue, drawn from the $16^{\mathrm{th}}$ and $84^{\mathrm{th}}$ percentile estimates).}
\label{fig:mass_mass_comp}
\end{figure}
\begin{figure*}
\centering 
\includegraphics[width=17.5cm,height=17.5cm,keepaspectratio,trim={0.5cm 4.5cm 0.5cm 4.5cm},clip]{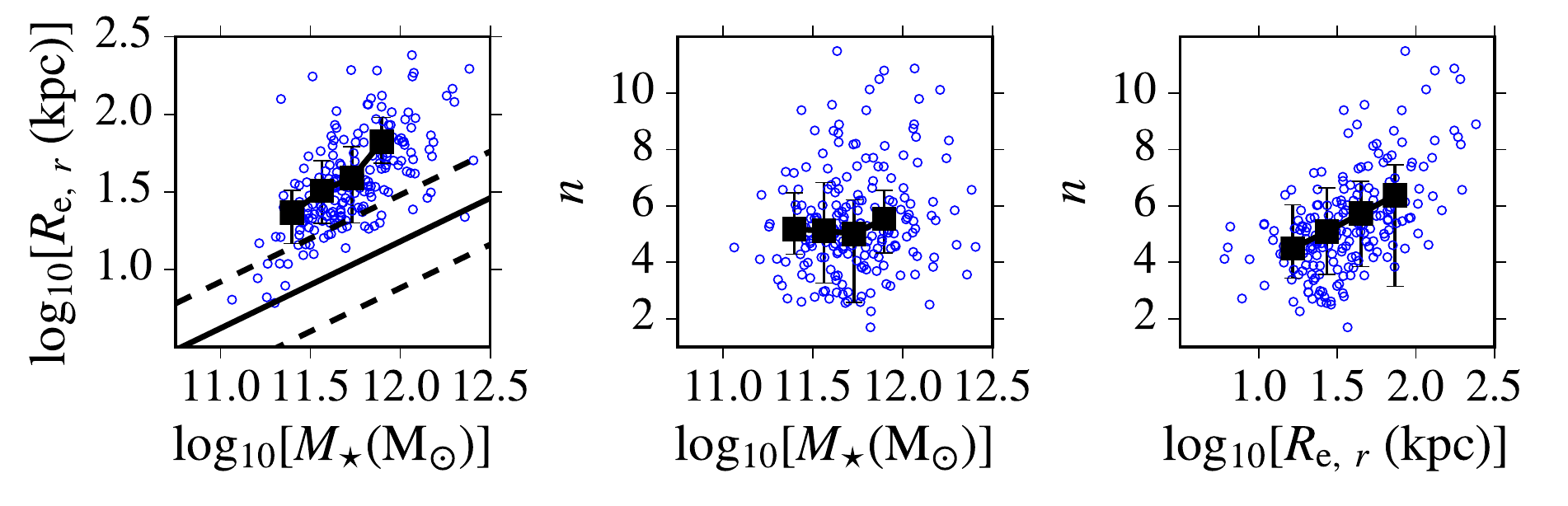} 
\caption{Correlations between the BCG structural parameters. For clarity, the data have been binned using a Scott's-rule optimised bin width to illustrate structure (omitting bins with fewer than 15 objects). The errorbars mark the $16^{\mathrm{th}}$ and $84^{\mathrm{th}}$ percentiles of each bin. The solid line is the relation for SDSS `early-type' galaxies ($n > 2.5$) from \protect\cite{2003MNRAS.343..978S}, with the dashed lines marking the $1\sigma$ scatter.}
\label{fig:bcgparamcomp}
\end{figure*}
\par In order to ensure the quality of our fits, we set several criteria which a fit must meet in order to be included in our analysis (see \citealt{2017ApJ...835...22P}; \citealt{2015MNRAS.448.2530Z}). These criteria include: a cut of ${{\chi}_{\nu}}^{2} \geq 3$ to remove any BCGs with bad residuals, a cut of $R_{e} \geq 800"$, $\Delta{R_{e}}/R_{e} \geq 0.8$ (where $\Delta{R_{e}}$ is the statistical error in $R_{e}$ from \textsc{GALFIT}, see \citealt{2008ApJ...683..644S}) and finally, $n \geq 14$. These parameters remove most fits deemed to be physically unrealistic; in total, approximately $\sim 9\%$ of objects were affected. Including the prior cuts on environmental parameters described in section \ref{analysis} brings the total number of BCGs included in our primary analysis sample to 198. We have checked and found no significant changes to our conclusions by including the fits which did not meet the stated criteria; for the sake of only adding additional statistical noise, we decided to omit them. 
\par As the S\'{e}rsic fits to our objects produce integrated magnitudes, they can be used as a proxy for stellar mass. In this work, we adopt the updated scaling relation of \cite{2011MNRAS.418.1587T}, which was used to estimate stellar masses for galaxies in the GAMA survey. They demonstrate that stellar masses can be estimated within $\sim 0.1$ dex of the SED mass estimates quoted in the MPA-JHU catalogues (195 objects; http://wwwmpa.mpa-garching.mpg.de/SDSS/DR7/; see \citealt{2003MNRAS.343..978S} for the method used to derive stellar masses). \cite{2011MNRAS.418.1587T} also reported that using the $i$-band as a tracer of galaxy mass produced results of similar quality to using the common proxy of NIR flux. The scaling relation is derived from SED fits to GAMA galaxies using SDSS $ugriz$ photometry, using stellar population synthesis models from the library of \cite{2003MNRAS.344.1000B} with a \cite{2003PASP..115..763C} IMF.
\par From the rest-frame $g-i$ colour of a galaxy and its absolute magnitude $M_{i}$, the stellar mass is estimated via the following empirical relation: 
\begin{equation} \label{eq:massscaling}
\mathrm{log_{10}[M/M_{\odot}]} \ = \ 1.15 \ + \ 0.70(g - i) \ - \ 0.4M_{i} \ ,
\end{equation}
where $M_{i}$ are the k/dust corrected absolute magnitude values in the $i$ band, derived from the best-fitting S\'{e}rsic profile for a given BCG and $g-i$ is its restframe colour. We measure the $g-i$ colours of our objects through fixed 30 kpc apertures, which we do to reduce potential biases which may occur across fits (see section \ref{simulations}).
\par We correct our objects for extinction from Galactic dust using the standard maps of {\cite{1998ApJ...500..525S} (updated normalisation in \citealt{2011ApJ...737..103S}), with a \cite{1999PASP..111...63F} reddening law. Typically, this correction is small; the mean values in each band are 0.08, 0.06 and 0.04 magnitudes in $g$, $r$ and $i$ respectively. We also apply a k-correction to our objects after correcting for dust. The k-corrections are based on the work of \cite{2010MNRAS.405.1409C} and work well for galaxies up to $z \sim 0.5$ in lieu of multi-band photometry. They are approximated by polynomials (e.g. \citealt{1998MNRAS.297..128C}), requiring only a redshift and input color. We use our aperture $g-r$ colors; the average k-corrections are  0.59, 0.19 and 0.10 magnitudes in $g$, $r$ and $i$ respectively. 
\par We caution the reader that our mass `errors' are drawn from the systematic errors output by the best-fit model from \textsc{GALFIT}; they are therefore likely to be underestimated. Instead, we offer a comparison between the stellar masses of our final sample of objects between a common subsample of 192 with valid masses drawn from the MPA-JHU value added catalogue in figure \ref{fig:mass_mass_comp} (\citealt{2003MNRAS.344.1000B} SP, \citealt{2001MNRAS.322..231K} IMF). The agreement is good, with the average scatter (approximately $\pm 0.06$ dex) similar to that predicted by \cite{2011MNRAS.418.1587T}. The small, positive offset is predominantly due to our choice of using model magnitude when computing our stellar masses, so as to account for additional mass in the wings of our BCGs. The BCGs span a large range in mass, from $11.0 \ < {\mathrm{log_{10}}}[M_{\star} (\mathrm{M_{\odot}})] < \ 12.5$, peaking at ${\mathrm{log_{10}}}[M_{\star} (\mathrm{M_{\odot}})] = 11.5$. 
\par As discussed at length in the previous section, we use the $r$-band output parameters to characterise our BCGs. There is, however, little variation in the two as both bands pick up light from predominantly the same stellar populations (e.g. \citealt{2016MNRAS.460.3458K}; \citealt{2007ApJ...659..162T}). We display comparisons between morphological parameter outputs for $g$/$i$ with $r$ in figure \ref{fig:re_n_comp} to illustrate this point. Moreover, as shown in figure \ref{fig:simi}, the overall scatter in the $i$-band output at $17^{\mathrm{th}}$ magnitude or brighter compared to input is reasonably small (e.g. ${\Delta}m_{i} \sim 0.2$ at $m_{i} = 17$); this result is also likely to be a worst-case scenario, given that the peak output $i$-band magnitude for our BCGs is $\sim 0.6$ magnitudes brighter than for the $r$-band.
\subsection{The Influence of Stellar Mass on BCG Structure}
\begin{figure*}
\centering 
\includegraphics[width=17.5cm,height=17.5cm,keepaspectratio,trim={1.0cm 0cm 1.5cm 0cm},clip]{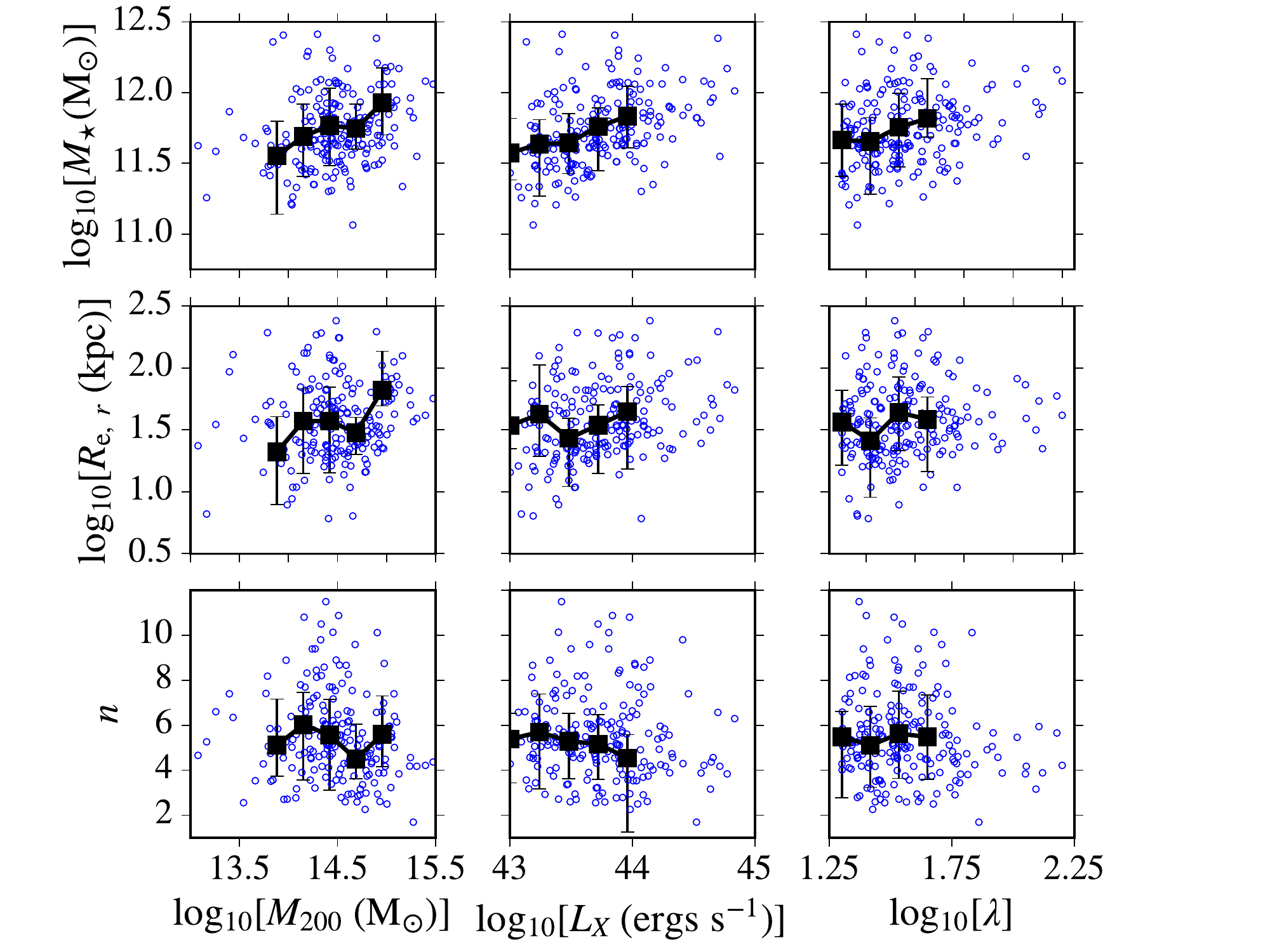} 
\caption{Correlations between the environmental parameters outlined in this study with BCG structural parameters and stellar mass. Bins are shown with $N$ $\geq$ 15 objects. The correlation between BCG mass and the properties of the host cluster appear more compelling than with the structural parameters, where we observe few significant correlations (see table \ref{t:full-spearman} and partial Spearman analysis in Appendix B).}
\label{fig:envbcg}
\end{figure*}
\par To determine which parameters are the primary drivers behind the correlations in our data, we have performed a Spearman rank analysis on our main sample (table \ref{t:full-spearman}). We have also provided a partial Spearman analysis as well to test the robustness of our results against selection effects, which is summarised in the Appendix for the parameters of interest (tables \ref{t:held-re}-\ref{t:held-z}; computed using \textsc{MATLAB}'s `\textsc{partialcorr}' routine), following a similar practice to \cite{1998MNRAS.297..128C}. We hold our significance level at the standard value of $p \leq 0.05$ throughout ($\mathrm{log_{10}}[p_{s}] \leq -1.301$). 
\begin{table*}
\centering
\begin{tabular}{l|lllllll}
 &$R_{e}$&$L_{X}$&$M_{200}$&$n$&$M_{\star}$&$\lambda$&$z$\\
$R_{e}$&-&0.26711&0.11014&0.58263&0.64668&0.12535&0.30716\\
$L_{X}$&-3.8404&$-$&0.43005&-0.14079&0.39447&0.54933&0.57387\\
$M_{200}$&-0.91578&-9.4719&$-$&-0.14423&0.26984&0.51566&0.29608\\
$n$&<-45&-1.3244&-1.3749&$-$&0.1548&-0.10605&-0.059135\\
$M_{\star}$&<-45&-7.9602&-3.911&-1.536&$-$&0.24395&0.50075\\
$\lambda$&-1.1095&-16.356&-14.187&-0.86639&-3.2872&$-$&0.30286\\
$z$&-4.9478&<-45&-4.6253&-0.39105&<-45&-4.8622&$-$\\
\end{tabular}
\caption{Full Spearman rank analysis for all the variables examined in this study for the $r$-band morphological parameters. The top half of the table lists the Spearman rank correlation coefficient ($r_{s}$), whereas the bottom half of the table provides the log of its corresponding p-value ($\mathrm{log_{10}}[p_{s}]$).}
\label{t:full-spearman}
\end{table*}
\par Figure \ref{fig:bcgparamcomp} suggests a strong correlation between the effective radius and stellar mass (luminosity) of our objects (Spearman rank coefficient, $r_{s}$ = 0.65, $\mathrm{log_{10}}[p_{s}] < -45$), which remains largely unchanged when any dependence on the environmental parameters are removed (see tables \ref{t:held-lx}, \ref{t:held-m200} and \ref{t:held-lambda}). Indeed, this behaviour is also seen in figure \ref{fig:envbcg}, which we discuss in detail in the upcoming section. These results indicate that the mass-size relation for BCGs seen here largely appears to exist independently of environment; this was also concluded by \cite{2015MNRAS.453.4444Z}. These observations support the scenario proposed by \cite{2014MNRAS.439.3189S}, who provided a semi-analytic model of BCG evolution since $z \sim 0.3$ and found no major environmental dependence on the sizes of early-type central galaxies (see section \ref{sampleselection}). Indeed, numerous other studies have found this to be true of cluster galaxies in general at $z < 1$. For example, \cite{2015MNRAS.450.1246K} modelled galaxies in a similar way to this work in the ESO distant cluster survey (EDisCS, \citealt{2005A&A...444..365W}), a HST survey of 20 cluster fields from 0.4 < $z$ < 0.8. Splitting the sample into to cluster and field galaxies using a threshold in velocity dispersion, they found no differences in the sizes of galaxies inside or outside of the cluster. \cite{2013MNRAS.428.1715H} reported similar findings in their work, reporting a doubling of massive ellipticals (including BCGs) in size from $z \sim 1$ to present, but no environmental dependence on the mass-size relation. 
\par The corresponding $R_{e}-M_{\star}$ relation from \cite{2003MNRAS.343..978S} derived for a general population of early-type galaxies independent of environment (defined as $n > 2.5$) has also been included for comparison. Our BCGs lie significantly above this relation ($\sim 0.5$ dex). \cite{2015MNRAS.453.4444Z} came to a similar conclusion in their work, finding that most of their BCGs lay significantly above this relation, especially for BCGs which they classified morphologically as `cD' types. The more massive and extended nature of BCGs in comparison to elliptical non-BCGs has been also found by numerous other studies since $z \sim 1$ (\citealt{2007MNRAS.379..867V}, \citealt{2014ApJ...797...62V}, \citealt{2009MNRAS.395.1491B}), although some argue that, when matched in colour and mass, there are few differences between BCGs and their satellites (e.g. \citealt{2009MNRAS.398.1129G}). 
\begin{figure*}
\centering 
\includegraphics[width=17.5cm,height=17.5cm,keepaspectratio,trim={0cm 4cm 0cm 4cm},clip]{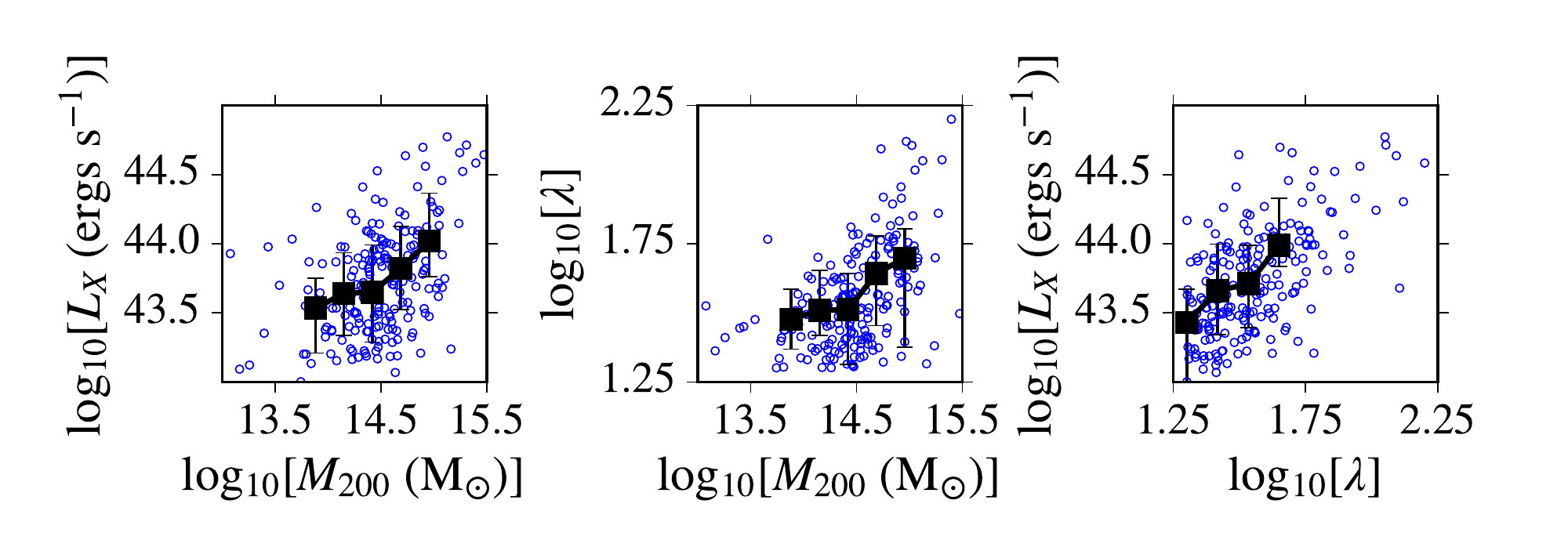} 
\caption{Correlations between characteristics of the host cluster (bins shown with $N \geq 15$ objects). The physical properties of clusters are highly correlated; hence the importance of accounting for selection.}
\label{fig:environmentcomp}
\end{figure*}
\par  Comparing with the correlation between mass and effective radius, we also find a weaker correlation between stellar mass and S\'{e}rsic index ($r_{s} = 0.1548$, $\mathrm{log_{10}}[p_{s}] = -1.536$). The consistency of the BCGs in our sample is likely due to the fact that they all have early-type morphologies and are bulge-dominated (sample median $n = 5.26 \pm 2.07$). Some studies have found correlations between $n$ and $M_{\star}$ (e.g. \citealt{2009MNRAS.398.1129G}); however, \cite{2015MNRAS.453.4444Z} reported that this relation is driven by the strong dependence of $n$ on $R_{e}$ than any separate dependence of mass on S\'{e}rsic index (rightmost panel of figure \ref{fig:bcgparamcomp}, $r_{s} = 0.58$, $\mathrm{log_{10}}[p_{s}] < -45$), finding that this relation disappeared when they chose to use SED-based masses (though their conclusions were otherwise unchanged when they used a scaling relation). They did however find a visual morphological dependence on S\'{e}rsic index when they split their sample into E and cD-type BCGs. Indeed, this weak, positive correlation is removed by accounting for any $R_{e}$ dependence through applying a partial Spearman (table \ref{t:held-re}).
\par The prominent anti-correlation between $n$ and $M_{\star}$ that then arises as a result of removing any dependence on $R_{e}$ ($r_s = -0.3641$, $\mathrm{log_{10}}[p_{s}] = -6.871$) could therefore be arise from two possible effects: a result of evolution, or underestimation of the slope due to surface-brightness effects which was also demonstrated by our simulations (figure \ref{fig:simr}). From our simulations in section \ref{simulations}, we believe it is more likely that the latter is the cause behind the observed scatter in this relation, due to some contamination with a number of profiles suffering from this effect. Indeed, when fixing for redshift, there is no anti-correlation present (table \ref{t:held-z}, $r_{s} = 0.2000$, $\mathrm{log_{10}}[p_{s}] = -2.321$). At the very least underestimating the slope cannot be ruled out as the source of the anti-correlation; nevertheless, although we observe an increased scatter in $R_{e}$ and magnitude at decreasing surface brightness, we see no evidence for any real change in direction of the bias (figure \ref{fig:simr}). It is worth noting that \cite{2011MNRAS.414..445S} also reported no evolutionary dependence on BCG profile slope between $0.25 < z < 1$ from their S\'{e}rsic fits; \cite{2011ApJ...726...69A} measured a change in size from $0 < z < 0.6$, but also no prominent change in profile slope. In contrast, \cite{2009MNRAS.395.1491B} did find that BCGs are more massive and extended than field or non-BCG satellite galaxies; however, \cite{2009MNRAS.394.1213W}, using SDSS data, found little difference. Discrepancies between results therefore appear to lie in the method of selection, the method of defining environment and whether to take a single or multiple-component approach when fitting. 
\subsection{How do the Characteristics of BCGs Relate to their Environment?}
We compare the properties of our BCGs with that of their host cluster environments in figure \ref{fig:envbcg}. It is immediately obvious that the masses of our BCGs are significantly correlated with all three environmental properties at the focus of this study; however, the strength of the correlation largely varies depending on the property of interest (see table \ref{t:full-spearman}). The stellar masses of our BCGs are the most strongly correlated with X-ray luminosity ($r_{s} = 0.394$, $\mathrm{log_{10}}[p_{s}] = -7.9602$) and are the least correlated with richness ($r_{s} = 0.244$, $\mathrm{log_{10}}[p_{s}] = -3.287$). \cite{2015MNRAS.453.4444Z} found a similar result with their BCGs, in that they measured a correlation between BCG stellar mass and cluster dynamical mass, albeit with large scatter (Pearson rank coefficient = 0.17, see paper for details). They provided arguments against the correlation being caused due to more massive haloes being populated by more massive BCGs simply by chance, due to the dominance of BCGs in comparison to the general population of cluster galaxies (e.g., \citealt{2007MNRAS.379..867V}).
\par All three environmental parameters, although measured independently, suffer from some degeneracy (more massive haloes are generally more likely to be occupied by a larger number of galaxies and contain a larger amount of bound ICM, e.g., \citealt{2016MNRAS.463.1929M}, \citealt{2011ApJ...728..126W}, \citealt{2010ApJ...709...97L}); the relevant environmental correlations are shown for reference in figure \ref{fig:environmentcomp}. To address potential selection biases which may arise with redshift (e.g. figure \ref{fig:lxz}), we test our correlations independently of redshift (table \ref{t:held-z}) to analyse the robustness of our results. When doing so, we find our correlations remain (albeit at reduced strength) for both X-ray luminosity ($r_{s} = 0.16$, $\mathrm{log_{10}}[p_{s}] = -1.61$) and cluster mass ($r_{s} = 0.14$, $\mathrm{log_{10}}[p_{s}] = -1.37$), but not for richness, which drops below significance ($r_{s} = 0.1$, $\mathrm{log_{10}}[p_{s}] = -0.8$). We discuss our interpretation of the lack of correlation with richness below. 
\par As shown in figure \ref{fig:environmentcomp}, the X-ray luminosity of a cluster is clearly dependent on mass and is often used as an alternative proxy for the former. However, depending on the dynamical state of the cluster, such measurements are prone to their own biases (e.g.  \citealt{2007ApJ...655...98N}). Relaxed, highly-evolved clusters have been found generally to be more likely to host an X-ray luminous cool core than clusters out of dynamical relaxation; indeed, the degree of offset of the BCG correlates inversely with the X-ray luminosity of a cluster (e.g. \citealt{2009MNRAS.398.1698S}, \citealt{2012MNRAS.422.2213S}). It would therefore follow that one of the drivers behind the strength of this relation is the tendency of more massive BCGs to be located within clusters with a greater degree of dynamical relaxation and structural evolution, potentially where the degree of `dominance' of the BCG is large (e.g., \citealt{2003MNRAS.343..627J}). Of course, there are physical mechanisms that add further complications to this assumption; an increased abundance of radio-loud AGN have been found in BCGs residing in cool core clusters (e.g. \citealt{1990AJ.....99...14B}, \citealt{1999MNRAS.306..857C}), the feedback from which are thought to be capable of heating the ICM (e.g. \citealt{2014ApJ...785...44M}, \citealt{2014ApJ...784...78R}, \citealt{2007MNRAS.379..894B}).
\par There is, as mentioned in section \ref{stellarmass}, little evidence for any independent environmental dependence on the scale sizes of our BCGs from mass. Although there is an apparent correlation with X-ray luminosity on effective radius ($r_s = 0.26711$, $\mathrm{log_{10}}[p_{s}] = -3.8404$), it entirely disappears at fixed stellar mass ($r_s = 0.028219$, $\mathrm{log_{10}}[p_{s}] = -0.1592$); suggesting that stellar mass rather than effective radius is the main driver behind the observed correlation. A similar conclusion was reached by \cite{2015MNRAS.453.4444Z} who also found, after visually classifying their BCGs into E and cD types (`bulge only' versus `bulge+envelope', see \citealt{2015MNRAS.448.2530Z} for the classification method), that cD types constitute a significantly more massive and extended population than E-types. \cite{2015MNRAS.453.4444Z} also reported a weak environmental dependence on their visual morphologies, with cD-type BCGs generally inhabiting marginally more massive, denser haloes than E-type BCGs. A larger fraction of BCGs with cD-type haloes were found by \cite{2005MNRAS.364.1354B} to reside in more X-ray luminous clusters. As we do not visually classify our BCGs due to the fact we are unlikely to possess the necessary photometric depth, we cannot provide a direct comparison; nevertheless, it would be interesting to explore large epochs of cosmic history to determine if this morphological dependence holds at higher redshift. Next-generation surveys, such as the Large Synoptic Survey Telescope (LSST) which constitute both large volumes and deep photometry may be the key for solving such problems (e.g., \citealt{2008arXiv0805.2366I}). 
\par As aforementioned, cluster richness, of the properties featured in this work, is the least significantly correlated with BCG properties, with no significant correlation present with stellar mass independently of redshift. This result may arise because it represents a weaker means of quantifying environment, in that it is simply a proxy for the number of galaxies attributed to a cluster. A richness value provides minimal information about the nature of the cluster galaxies; the influence of several neighbours of comparable mass to a BCG would likely have a larger influence than an equal number of much smaller neighbours, nevertheless the richness estimators in each case would be equivalent (e.g. \citealt{1970ApJ...162L.149B} provided a basic classification scheme for galaxy clusters with this issue in mind). However, the richness values can be used to provide a robust measure of the total stellar mass of a cluster when coupled with abundance-matching, as demonstrated by \cite{2015MNRAS.449.1897O}.
\begin{figure*}
\centering 
\includegraphics[width=17cm,height=17cm,keepaspectratio,trim={0.2cm 4.5cm 0.5cm 4cm},clip]{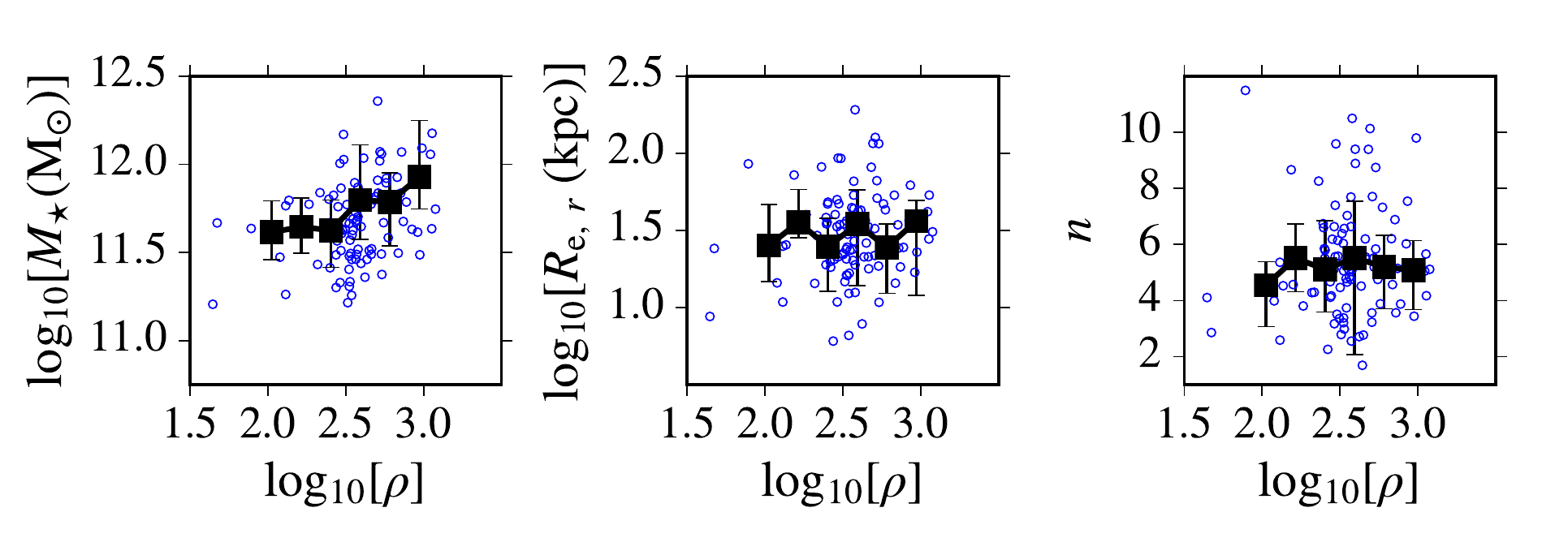}  
\caption{Correlations between luminosity-weighted environmental density, $\mathrm{log_{10}}[\rho]$, with BCG properties (bins shown with $N$ $\geq$ 5 objects) for 102 BCGs in common with the catalogues of \protect\cite{2012A&A...540A.106T}. As before, there is a correlation with mass, but no correlation with the structural parameters.}
\label{fig:densitymorph}
\end{figure*}
\par Arguably, a more physically-motivated measure of environment for our purposes than richness would take overall luminosities of galaxies within a cluster into account; this point was also raised by \cite{2015MNRAS.448.2530Z}. Such a measure is shown in figure \ref{fig:densitymorph} using the luminosity-weighted environmental density maps of \cite{2012A&A...540A.106T}, smoothed to scales of $1 \ \mathrm{Mpc \ h^{-1}}$. The maps are computed from SDSS DR8 data and constitute the largest contiguous region of the SDSS footprint; we matched the SPIDERS BCGs with the catalogues within 3$"$, finding 102 common objects which had a corresponding measurement of environmental density. Although we cannot provide a complete comparison as we lack coverage for the full sample of objects, the result appears promising. In common with figure \ref{fig:envbcg}, there is a significant correlation between environmental density and BCG stellar mass ($r_{s} = 0.3233$, $\mathrm{log_{10}}[p_{s}] = -3.0128$), but no correlation between either effective radius or S\'{e}rsic index (Spearman rank correlation coefficients for $n$ and $R_{e}$ 0.1369 and 0.0117 respectively, with corresponding $\mathrm{log_{10}}[p_{s}]$ values of -0.6185 and -0.7701). Even after accounting for the strong $L_{X}$ dependence through our partial Spearman analysis, the correlation between BCG mass and environmental density remains significant ($r_{s} = 0.2554$, $\mathrm{log_{10}}[p_{s}] = -2.0023$), as does the $L_{X} - M_{\star}$ relation when controlling for environmental density ($r_{s} = 0.3150$, $\mathrm{log_{10}}[p_{s}] = -2.8811$). This result suggests that our sample contains a significant fraction of clusters which are mature systems (i.e. self-contained, virialised), having accumulated the majority of their stellar component up to $z \sim 0.3$. 
\section{Discussion}
\par At fixed stellar mass, we do not measure any significant anti-correlation between redshift and scale size for our sample (table \ref{t:held-mbcg}); this therefore suggests that there is little overall evolution in the scale size of our BCGs. Due to the large number of rich, high-mass clusters in our sample compared with group-level systems, it is likely that many of the clusters in our sample are nearing maturity; this is reflected by the relatively homogeneous properties of the BCGs observed here. Our findings here are consistent with \cite{2011MNRAS.414..445S}, who also used an X-ray selected cluster sample similar in redshift and cluster X-ray luminosity to the SPIDERS sample. It also appears that any environmental dependence on the size-stellar mass relation for BCGs is minimal for our sample at the redshift and halo mass range of this study (median $M_{200} = 1.4 \times 10^{14} \ \mathrm{M_{\odot}}$). Stellar mass, over the environmental properties featured in this study, arises as the more important factor governing BCG morphology. For example, \cite{2009MNRAS.398.1129G}, reporting a similar result in their work, interpreted the lack of an obvious $n - M_{200}$ relation as indicating that there is no clear mass threshold where a dark matter halo is capable of producing spheroidal centrals. However, they found a strong trend between stellar mass and S\'{e}rsic index. This claim was disputed by \cite{2015MNRAS.448.2530Z}, who found no such trend; however, they also found little environmental dependence on the structural parameters of BCGs up to $z \sim 0.1$ when their sample was not split by visual morphology. 
\par Here, our findings suggest a trend between $n - M_{\star}$; though in general, we make the simplifying assumption due to the higher peak redshift of our BCGs that they are single component objects, and therefore do not attempt to fit a bulge+disc. As we have demonstrated in our simulations in section \ref{simulations}, $n = 1$, `disk-like' profiles demonstrate unpredictable behaviour more rapidly than `bulge-like', higher-$n$ profiles. It would therefore, at the magnitude range of the BCGs in this study, be difficult to draw any meaningful conclusions from fitting a dual component profile (e.g. S\'{e}rsic+exponential) for any but the most luminous galaxies in our study. As wide surveys with significantly deeper photometry become readily available over the next decade, they would be an ideal testbed at higher redshift to examine the observed dual-component nature of some BCGs seen at $z < 0.1$ (e.g. \citealt{2017arXiv170701904H}), and, by extension, the build-up of the ICL.
\par It follows that, if the BCGs in our sample display little morphological dependence with environment, any influence of environment on their evolution must have been apparent at an earlier point in the assembly process. Due to the fact that our BCGs are more massive and extended than the general population of $n > 2.5$ elliptical galaxies (section \ref{results}), the cluster potential well must have some influence in the past on the formation of BCGs. For example, the work of \cite{2007MNRAS.379..867V} found BCGs to have a higher dynamical-to-stellar mass ratio, indicating that they contained a larger fraction of dark matter compared to a sample of colour-matched non-BCG galaxies taken from the SDSS. 
\par Various studies have predicted the growth in stellar mass of BCGs, with a wide range of results predicting BCGs doubling in size since $z \sim 0.3$ to predicting size growth of less than 20\% since $z \sim 1$ (e.g., \citealt{2009MNRAS.395.1491B}, \citealt{2014ApJ...797...62V}, \citealt{2011ApJ...726...69A}, \citealt{2011MNRAS.414..445S}). The discrepancies lie not only in the method of measurement (e.g., single profile-modelling versus dual-profile modelling, to account for the stellar halo seen in BCGs), but also the method of sample selection; an early study, \cite{2003Ap&SS.285...51C}, argued that the growth rates seen in X-ray luminous clusters are modest since $z \sim 1$, with larger rates present in clusters with low X-ray luminosities. The early build-up of stellar mass in BCGs (e.g., \citealt{2009Natur.458..603C}) as well as observations of an established red sequence in clusters at $z > 1.5$ (e.g., \citealt{2016ApJ...816...83C}) still present a challenge to simulations, some of which predict a large mass increase in BCGs between $0 < z < 1$ (e.g., \citealt{2007MNRAS.375....2D} predicted a fourfold increase during this timescale). An improved understanding of the formation of the ICL, such as the stripping of stars from central galaxies during the cluster assembly process, is therefore required to understand the ongoing discrepancies between simulations and observations of BCGs (e.g., \citealt{2012MNRAS.425.2058B}, \citealt{2015MNRAS.449.2353B}).
\section{Conclusions}
We created a sample of 329 BCGs from the X-ray selected SPIDERS clusters survey, and investigated three cluster properties of interest: X-ray luminosity, richness, and halo mass, the last property of which we estimated through the cluster velocity dispersion. We modelled our BCGs with single-S\'{e}rsic profiles using the \textsc{SIGMA} pipeline. We tested the ability of our pipeline to recover parameters from SDSS data using $10^4$ model profiles, outlining a final science sample of 198 BCGs. Finding the results of our best-fitting parameters to be generally consistent across bands, we derived stellar masses for our BCGs based on \cite{2011MNRAS.418.1587T}. We conclude the following:
\begin{itemize}
\item Our simulations suggest a strong codependency between $n$, $R_{e}$ and apparent magnitude. 
\item We detected a negative bias in S\'{e}rsic index with effective radius, as a result of the degeneracy between the background level and profile wings. We also found a significant increase in the scatter of output effective radii at fainter magnitudes. This behaviour occurred regardless of band or input S\'{e}rsic index/effective radius. We used this information to approximate the fitting magnitude limit of our sample ($m_{r} = 16.5$).  
\item The scale sizes of the BCGs in our sample are highly correlated with their stellar masses, in common with numerous other studies (section \ref{results}). The BCGs are also significantly more massive and extended than the general $n > 2.5$ population of galaxies in the SDSS. 
\item There is a weak correlation between BCG mass and S\'{e}rsic index, inferring that more massive BCGs may tend towards having slightly more centrally-concentrated light profiles.
\item Significant correlations exist between the masses of our BCGs and all three of the cluster properties explored in this study (richness, cluster mass and X-ray luminosity). However, fixing for redshift, we do not find any significant correlation with richness - indicating that it is likely to be less useful measure of environment in this context.
\item There is no evidence that environment, at the redshift of our clusters, has any influence over the size-stellar mass relation of our BCGs, nor is there evidence for any correlation between the profile slopes of our BCGs and the cluster environment.
\item For a reduced sample of 102 BCGs, the environmental density is highly correlated with stellar mass, but no correlation is present with either structural parameter. A partial Spearman analysis reveals this correlation to be largely independent of X-ray luminosity.
\end{itemize}
\par The picture is therefore that BCGs, in rich, X-ray selected clusters appear to have no significant environmental dependence on their structures, independently of their mass, up to $z \sim 0.3$. If the primary driver behind growth of BCGs is indeed through multiple mergers (e.g. \citealt{1975ApJ...202L.113O}), it is likely that within the $M_{200}$ range of the clusters explored here, the mass assembly has predominantly occurred at earlier times, with growth slowing due to the large dynamical friction timescales associated with massive clusters at late times. Our work supports the scenario of the homogeneity presented by BCGs in massive clusters up to intermediate redshifts (e.g. \citealt{1998MNRAS.297..128C}, \citealt{2009Natur.458..603C}, \citealt{2008MNRAS.387.1253W}, \citealt{2011MNRAS.414..445S}). The full catalogue of objects is published electronically; a description of catalogue parameters is included in Appendix C.
\section*{Acknowledgements}
We thank the anonymous referee for their insightful comments that greatly improved the clarity of the paper. K.E. Furnell is supported by a Science and Technologies Funding Council (STFC) award. C.A. Collins, I.K. Baldry and L.S. Kelvin are supported by an STFC research grant (ST/M000966/1). K.E. Furnell would also like to personally thank the SPIDERS and CODEX teams for their cooperation, collaboration and support throughout this work. K.E. Furnell would also like to thank L.S. Kelvin for his consistent advice, support and enabling the use of the \textsc{SIGMA} software throughout this project.
\par This paper makes use of data from the Sloan Digital Sky Survey. Funding for the Sloan Digital Sky Survey IV has been provided by the Alfred P. Sloan Foundation, the U.S. Department of Energy Office of Science, and the Participating Institutions. SDSS acknowledges support and resources from the Center for High-Performance Computing at the University of Utah. The SDSS web site is www.sdss.org.
\par SDSS is managed by the Astrophysical Research Consortium for the Participating Institutions of the SDSS Collaboration including the Brazilian Participation Group, the Carnegie Institution for Science, Carnegie Mellon University, the Chilean Participation Group, the French Participation Group, Harvard-Smithsonian Center for Astrophysics, Instituto de Astrof\'{i}sica de Canarias, The Johns Hopkins University, Kavli Institute for the Physics and Mathematics of the Universe (IPMU) / University of Tokyo, Lawrence Berkeley National Laboratory, Leibniz Institut f\"{u}r Astrophysik Potsdam (AIP), Max-Planck-Institut f\"{u}r Astronomie (MPIA Heidelberg), Max-Planck-Institut f\"{u}r Astrophysik (MPA Garching), Max-Planck-Institut f\"{u}r Extraterrestrische Physik (MPE), National Astronomical Observatories of China, New Mexico State University, New York University, University of Notre Dame, Observat\'{o}rio Nacional / MCTI, The Ohio State University, Pennsylvania State University, Shanghai Astronomical Observatory, United Kingdom Participation Group, Universidad Nacional Aut\'{o}noma de M\'{e}xico, University of Arizona, University of Colorado Boulder, University of Oxford, University of Portsmouth, University of Utah, University of Virginia, University of Washington, University of Wisconsin, Vanderbilt University, and Yale University.
\bibliographystyle{mnras}
\bibliography{bibliography} 
\bsp

\clearpage
\section*{Appendix A - Simulated Profile Outputs for $g$/$i$} 

\noindent\begin{minipage}{\textwidth}
    \centering
    \includegraphics[width=21cm,height=21cm,keepaspectratio,keepaspectratio,trim={1.5cm 0cm 1cm 0cm},clip]{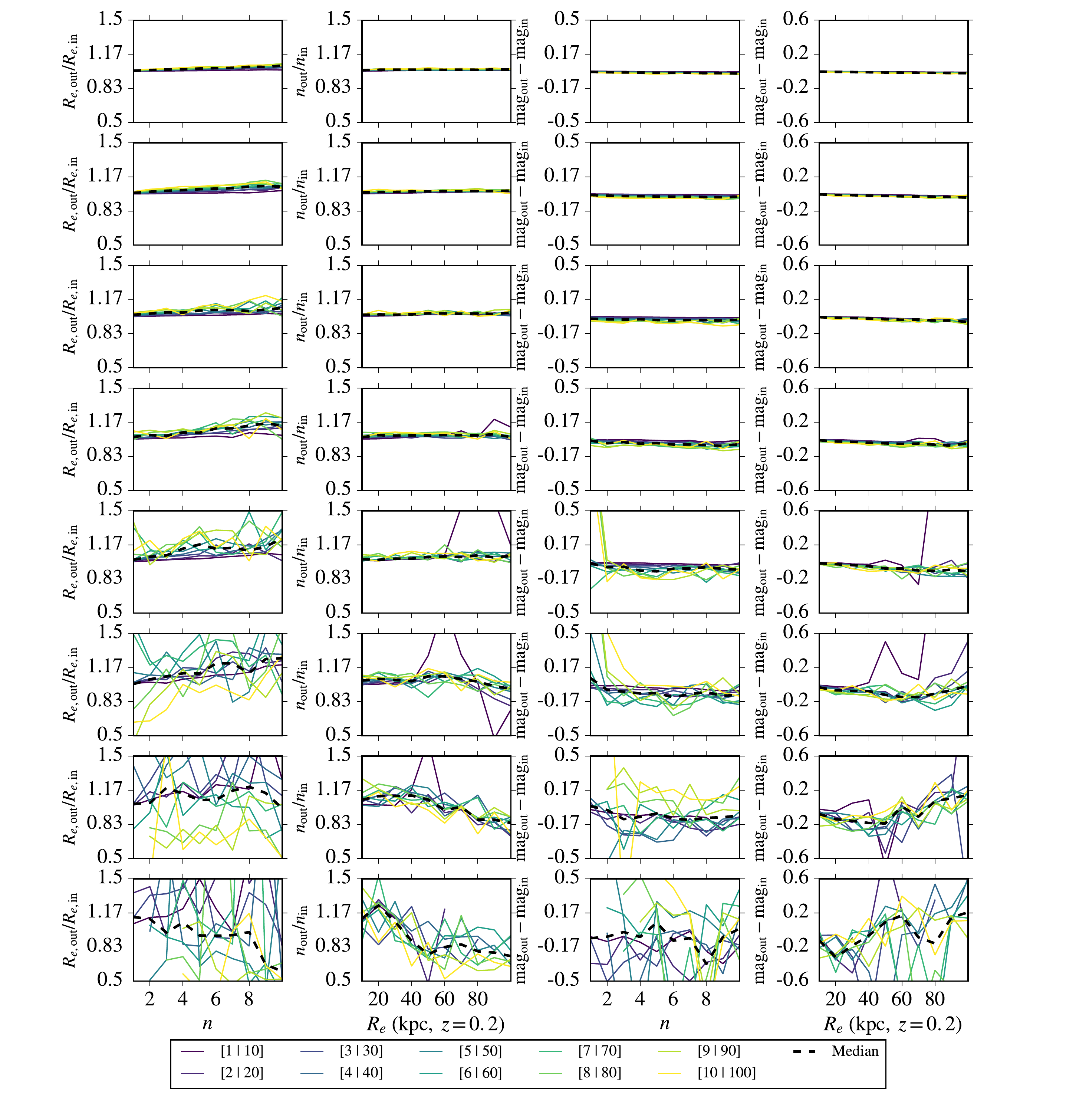}
    \captionof{figure}{As in figure \ref{fig:simr}, but for the $g$-band.}
    \label{fig:simg}
\end{minipage}


\begin{figure*}
\centering
\includegraphics[width=21cm,height=21cm,keepaspectratio,keepaspectratio,trim={1.5cm 0cm 1cm 0cm},clip]{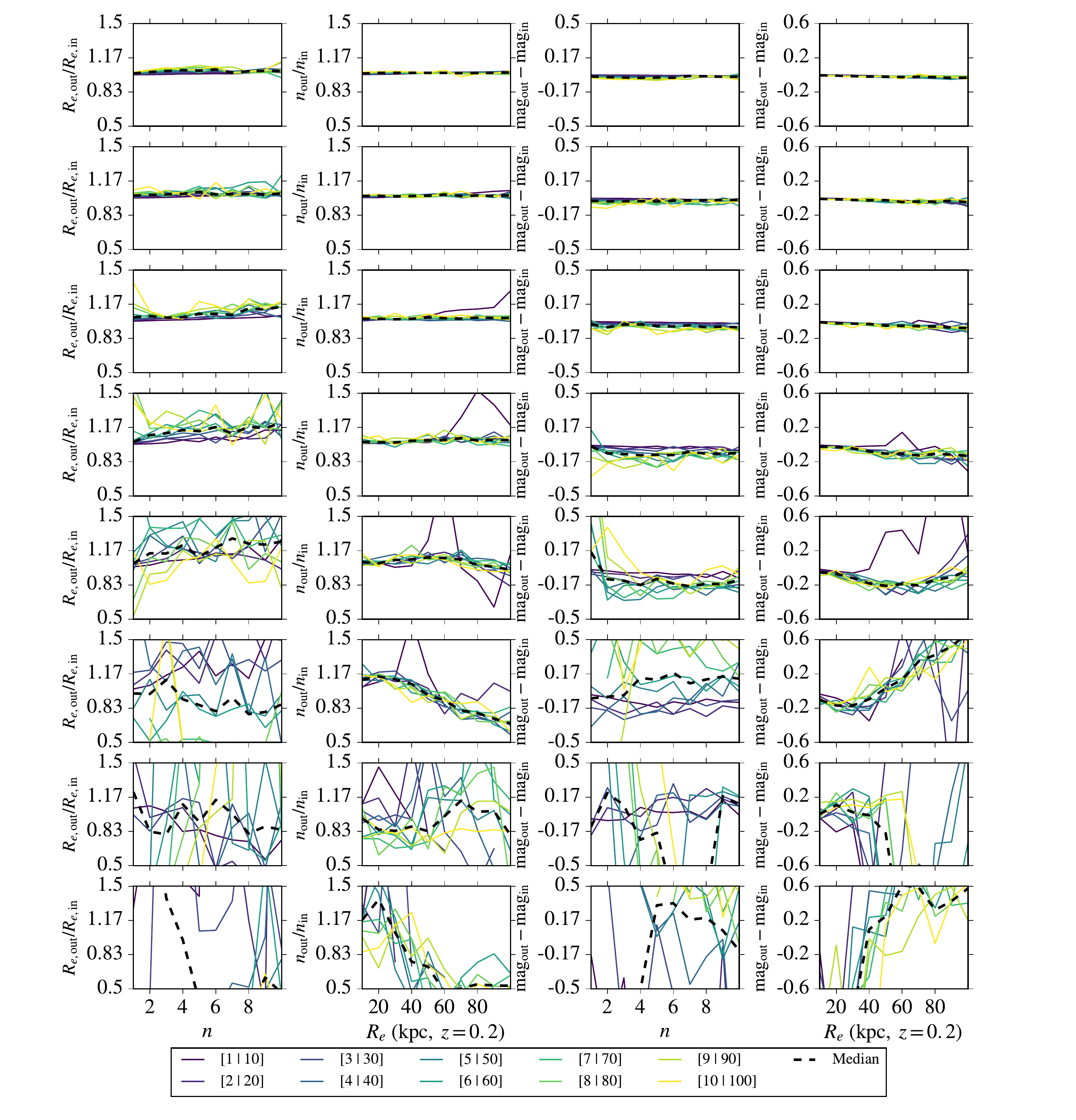}
\caption{As in figure \ref{fig:simr}, but for the $i$-band.}
\label{fig:simi}
\end{figure*}

\clearpage

\section*{Appendix B - Partial Spearman Analysis}
\noindent\begin{minipage}{\textwidth}
\centering
\captionof{table}{As in table \ref{t:full-spearman}, but for fixed $R_{e}$.}
\begin{tabular}{l|lllllll}
 &$R_{e}$&$L_{X}$&$M_{200}$&$n$&$M_{\star}$&$\lambda$&$z$ \\
$R_{e}$&$-$&$-$&$-$&$-$&$-$&$-$&$-$ \\
$L_{X}$&$-$&$-$&0.41439&-0.37348&0.3074&0.54307&0.53253 \\
$M_{200}$&$-$&-8.8905&$-$&-0.27916&0.25043&0.48343&0.24611 \\
$n$&$-$&-7.2236&-4.1668&$-$&-0.36407&-0.22452&-0.30248 \\
$M_{\star}$&$-$&-4.9761&-3.4282&-6.871&$-$&0.21417&0.42349 \\
$\lambda$&$-$&-15.858&-12.265&-2.8318&-2.6114&$-$&0.28225 \\
$z$&$-$&-15.17&-3.3243&-4.8288&-9.2931&-4.2512&$-$ \\
\end{tabular}
\label{t:held-re}
\end{minipage}\newline

\noindent\begin{minipage}{\textwidth}
\centering
\captionof{table}{As in table \ref{t:full-spearman}, but for fixed $L_{X}$.}
\begin{tabular}{l|lllllll}
 &$R_{e}$&$L_{X}$&$M_{200}$&$n$&$M_{\star}$&$\lambda$&$z$ \\
$R_{e}$&$-$&$-$&-0.022206&0.64945&0.60292&-0.042913&0.20203 \\
$L_{X}$&$-$&$-$&$-$&$-$&$-$&$-$&$-$ \\
$M_{200}$&-0.12139&$-$&$-$&-0.12644&0.099939&0.33868&0.027848 \\
$n$&-24.376&$-$&-1.1198&$-$&0.22026&-0.049023&0.032956 \\
$M_{\star}$&-20.257&$-$&-0.7925&-2.7399&$-$&0.021243&0.37383 \\
$\lambda$&-0.26097&$-$&-5.9754&-0.30733&-0.11553&$-$&-0.018234 \\
$z$&-2.3652&$-$&-0.15681&-0.19054&-7.2367&-0.097594&$-$
\end{tabular}
\label{t:held-lx}
\end{minipage}\newline

\noindent\begin{minipage}{\textwidth}
\centering
\captionof{table}{As in table \ref{t:full-spearman}, but for fixed $M_{200}$.}
\begin{tabular}{l|lllllll}
 &$R_{e}$&$L_{X}$&$M_{200}$&$n$&$M_{\star}$&$\lambda$&$z$ \\
$R_{e}$&$-$&0.26515&$-$&0.61064&0.64182&0.082771&0.30621 \\
$L_{X}$&-3.7962&$-$&$-$&-0.063337&0.34106&0.43534&0.5292 \\
$M_{200}$&$-$&$-$&$-$&$-$&$-$&$-$&$-$ \\
$n$&-20.894&-0.42555&$-$&$-$&0.19977&-0.033542&-0.0023672 \\
$M_{\star}$&-23.651&-6.056&$-$&-2.3209&$-$&0.13708&0.47805 \\
$\lambda$&-0.60848&-9.8358&$-$&-0.19451&-1.2666&$-$&0.20791 \\
$z$&-4.9404&-14.958&$-$&-0.011621&-11.974&-2.4828&$-$ 
\end{tabular}
\label{t:held-m200}
\end{minipage}\newline

\noindent\begin{minipage}{\textwidth}
\centering
\captionof{table}{As in table \ref{t:full-spearman}, but for fixed $n$.}
\begin{tabular}{l|lllllll}
 &$R_{e}$&$L_{X}$&$M_{200}$&$n$&$M_{\star}$&$\lambda$&$z$ \\
$R_{e}$&$-$&0.44098&0.24644&$-$&0.69269&0.22933&0.42767 \\
$L_{X}$&-10.102&$-$&0.41151&$-$&0.4339&0.54493&0.57356 \\
$M_{200}$&-3.3321&-8.766&$-$&$-$&0.287&0.48069&0.26014 \\
$n$&$-$&$-$&$-$&$-$&$-$&$-$&$-$ \\
$M_{\star}$&-28.904&-9.7689&-4.3829&$-$&$-$&0.26129&0.52637 \\
$\lambda$&-2.9377&-15.981&-12.116&$-$&-3.6976&$-$&0.30112 \\
$z$&-9.4822&-17.987&-3.6684&$-$&-14.78&-4.7886&$-$ 
\end{tabular}
\label{t:held-n}
\end{minipage}\newline

\noindent\begin{minipage}{\textwidth}
\centering
\captionof{table}{As in table \ref{t:full-spearman}, but for fixed $M_{\star}$.}
\begin{tabular}{l|lllllll} 
 &$R_{e}$&$L_{X}$&$M_{200}$&$n$&$M_{\star}$&$\lambda$&$z$ \\
$R_{e}$&$-$&0.028219&-0.086562&0.6433&$-$&-0.045577&-0.014766 \\
$L_{X}$&-0.1592&$-$&0.36228&-0.20772&$-$&0.51141&0.46609 \\
$M_{200}$&-0.64727&-6.8054&$-$&-0.21471&$-$&0.45626&0.16035 \\
$n$&-23.79&-2.4788&-2.6225&$-$&$-$&-0.15252&-0.14405 \\
$M_{\star}$&$-$&$-$&$-$&$-$&$-$&$-$&$-$ \\
$\lambda$&-0.28088&-13.863&-10.846&-1.4956&$-$&$-$&0.2169 \\
$z$&-0.077573&-11.345&-1.6193&-1.3676&$-$&-2.6685&$-$ \\

\end{tabular}
\label{t:held-mbcg}
\end{minipage}\newline

\begin{table*}
\centering
\caption{As in table \ref{t:full-spearman}, but for fixed $\lambda$.}
\begin{tabular}{l|lllllll}
 &$R_{e}$&$L_{X}$&$M_{200}$&$n$&$M_{\star}$&$\lambda$&$z$ \\
$R_{e}$&$-$&0.25929&0.04714&0.60374&0.63716&$-$&0.29982 \\
$L_{X}$&-3.6471&$-$&0.21151&-0.080153&0.33863&$-$&0.51151 \\
$M_{200}$&-0.29278&-2.5564&$-$&-0.13082&0.16229&$-$&0.13857 \\
$n$&-20.324&-0.58229&-1.1791&$-$&0.18071&$-$&-0.0135 \\
$M_{\star}$&-23.219&-5.9738&-1.6507&-1.9649&$-$&$-$&0.47589 \\
$\lambda$&$-$&$-$&$-$&$-$&$-$&$-$&$-$ \\
$z$&-4.7505&-13.87&-1.2878&-0.070441&-11.859&$-$&$-$ 
\end{tabular}
\label{t:held-lambda}
\end{table*}

\begin{table*}
\centering
\caption{As in table \ref{t:full-spearman}, but for fixed $z$.}
\begin{tabular}{l|lllllll}
 &$R_{e}$&$L_{X}$&$M_{200}$&$n$&$M_{\star}$&$\lambda$&$z$ \\
$R_{e}$&$-$&0.1256&0.016817&0.63101&0.58866&0.025858&$-$ \\
$L_{X}$&-1.1086&$-$&0.34469&-0.12338&0.15958&0.48296&$-$ \\
$M_{200}$&-0.089331&-6.1803&$-$&-0.16127&0.14449&0.44543&$-$ \\
$n$&-22.66&-1.0792&-1.6342&$-$&0.20006&-0.10169&$-$ \\
$M_{\star}$&-19.126&-1.6069&-1.3741&-2.3266&$-$&0.10323&$-$ \\
$\lambda$&-0.14409&-12.24&-10.315&-0.8125&-0.83019&$-$&$-$ \\
$z$&$-$&$-$&$-$&$-$&$-$&$-$&$-$
\end{tabular}
\label{t:held-z}
\end{table*}
\newpage

\clearpage

\section*{Appendix C - Table Schema}
\noindent\begin{minipage}{\textwidth}
\centering
\captionof{table}{Column descriptions for the parameters used in this study (table published electronically).}
\begin{tabular}{lll}
\hline
\multicolumn{1}{l|}{Column Name}                                    & \multicolumn{1}{l|}{Description}                                                                                                           & Units                    \\ \hline
\multicolumn{1}{l|}{SPIDERS\_ID}                                    & \multicolumn{1}{l|}{Unique Cluster ID from SPIDERS}                                                                                        & $-$                      \\
\multicolumn{1}{l|}{RA}                                             & \multicolumn{1}{l|}{Right ascension of BCG}                                                                                                & degrees                  \\
\multicolumn{1}{l|}{DEC}                                            & \multicolumn{1}{l|}{Declination of BCG}                                                                                                    & degrees                  \\
\multicolumn{1}{l|}{$z_{\mathrm{BCG, \ spec}}$}                     & \multicolumn{1}{l|}{Spectroscopic redshift of BCG}                                                                                         & $-$                      \\
\multicolumn{1}{l|}{$z_{\mathrm{BCG, \ spec}}\_\mathrm{err}$}       & \multicolumn{1}{l|}{Error on spectroscopic redshift of BCG}                                                                                & $-$                      \\
\multicolumn{1}{l|}{$z_{\mathrm{RM}}$}                              & \multicolumn{1}{l|}{Photometric redshift of cluster red sequence from \textsc{redMaPPer}}                
& $-$                      \\
\multicolumn{1}{l|}{$z_{\mathrm{RM}}\_\mathrm{err}$}       & \multicolumn{1}{l|}{Error on photometric redshift of cluster red sequence from \textsc{redMaPPer}}        
& $-$                      \\
\multicolumn{1}{l|}{$z_{\mathrm{clus, \ spec}}$}                    & \multicolumn{1}{l|}{Spectroscopic cluster redshift, estimated from biweight of cluster members} 
& $-$                      \\
\multicolumn{1}{l|}{$z_{\mathrm{clus, \ spec}}\_\mathrm{err, \ u}$} & \multicolumn{1}{l|}{Upper error bound on spectroscopic cluster redshift}                       & $-$                      \\
\multicolumn{1}{l|}{$z_{\mathrm{clus, \ spec}}\_\mathrm{err, \ l}$} & \multicolumn{1}{l|}{Lower error bound on spectroscopic cluster redshift}                       & $-$                      \\
\multicolumn{1}{l|}{$\sigma_{200}$}                                 & \multicolumn{1}{l|}{Velocity dispersion of cluster}                                                                                        & $\mathrm{km \ s^{-1}}$   \\
\multicolumn{1}{l|}{$\sigma_{200}\_\mathrm{err, \ u}$}              & \multicolumn{1}{l|}{Upper error bound on velocity dispersion of cluster}                                                                   & $\mathrm{km \ s^{-1}}$   \\
\multicolumn{1}{l|}{$\sigma_{200}\_\mathrm{err, \ l}$}              & \multicolumn{1}{l|}{Lower error bound on velocity dispersion of cluster}                                                                   & $\mathrm{km \ s^{-1}}$   \\
\multicolumn{1}{l|}{$N$}                                            & \multicolumn{1}{l|}{Number of cluster members used when computing velocity dispersion}           
& $-$                      \\
\multicolumn{1}{l|}{$\lambda$}                                      & \multicolumn{1}{l|}{Cluster richness estimator from \textsc{redMaPPer}}                                                                             & $-$                      \\
\multicolumn{1}{l|}{$L_{X}/E_{z}$}                                  & \multicolumn{1}{l|}{Cluster X-ray luminosity from ROSAT}                                                                                   & $\mathrm{ergs \ s^{-1}}$ \\
\multicolumn{1}{l|}{$\mathrm{log_{10}}[M_{\mathrm{BCG}}]$}          & \multicolumn{1}{l|}{Stellar mass of BCG}                                                                                                   & $\mathrm{M_{\odot}}$     \\
\multicolumn{1}{l|}{$\mathrm{log_{10}}[M_{\mathrm{BCG}}]\_\mathrm{err}$}     & \multicolumn{1}{l|}{Error on stellar mass of BCG}                                                                                          & dex                      \\
\multicolumn{1}{l|}{$(g-i)$}     & \multicolumn{1}{l|}{$k$-corrected, dust corrected, $g-i$ colour of BCG, measured in a 30kpc aperture}                                                                                          & $-$                      \\
\multicolumn{1}{l|}{$(g-i)\_\mathrm{err}$}     & \multicolumn{1}{l|}{Error on $g-i$ colour of BCG}                                                                                          & $-$                      \\
\multicolumn{1}{l|}{$m_{g}$}                                        & \multicolumn{1}{l|}{$g$-band apparent magnitude from \textsc{SIGMA}}                                                                                & $-$                      \\
\multicolumn{1}{l|}{$m_{g}\_\mathrm{err}$}                                   & \multicolumn{1}{l|}{Error on $g$-band apparent magnitude from \textsc{SIGMA}}                                                                       & $-$                      \\
\multicolumn{1}{l|}{$R_{\mathrm{e}, \ g}$}                               & \multicolumn{1}{l|}{Effective radius of BCG from \textsc{SIGMA}, $g$-band}                                                                                    & arcseconds               \\
\multicolumn{1}{l|}{$R_{\mathrm{e}, \ g}\_\mathrm{err}$}                          & \multicolumn{1}{l|}{Error on effective radius of BCG from \textsc{SIGMA}, $g$-band}                                                                           & arcseconds               \\
\multicolumn{1}{l|}{$n_{g}$}                                            & \multicolumn{1}{l|}{S\'{e}rsic index of BCG from \textsc{SIGMA}, $g$-band}                                                                                  & $-$                      \\
\multicolumn{1}{l|}{$n_{g}\_\mathrm{err}$}                              & \multicolumn{1}{l|}{Error on S\'{e}rsic index of BCG from \textsc{SIGMA}, $g$-band}                                                                         & $-$                      \\
\multicolumn{1}{l|}{$b/a_{g}$}                                          & \multicolumn{1}{l|}{Axis ratio from \textsc{SIGMA}, $g$-band}                                                                                                 & $-$                      \\
\multicolumn{1}{l|}{$b/a_{g}\_\mathrm{err}$}                                          & \multicolumn{1}{l|}{Error on axis ratio from \textsc{SIGMA}, $g$-band}                                                                                                 & $-$                      \\
\multicolumn{1}{l|}{${\theta}_{g}$}                                          & \multicolumn{1}{l|}{Position angle from \textsc{SIGMA}, $g$-band}                                                                                                & degrees                      \\
\multicolumn{1}{l|}{${\theta}_{g}\_\mathrm{err}$}                                          & \multicolumn{1}{l|}{Error on position angle from \textsc{SIGMA}, $g$-band}                                                                                                 & degrees                      \\
\multicolumn{1}{l|}{${{\chi}_{\nu, \ g}}^2$}                                          & \multicolumn{1}{l|}{Reduced ${\chi}^2$, $g$-band}                                                                                                 & $-$                      \\
\multicolumn{1}{l|}{$m_{r}$}                                        & \multicolumn{1}{l|}{$r$-band apparent magnitude from \textsc{SIGMA}}                                                                                & $-$                      \\
\multicolumn{1}{l|}{$m_{r}\_\mathrm{err}$}                                   & \multicolumn{1}{l|}{Error on $r$-band apparent magnitude from \textsc{SIGMA}}                                                                       & $-$                      \\
\multicolumn{1}{l|}{$R_{\mathrm{e},\ r}$}                               & \multicolumn{1}{l|}{Effective radius of BCG from \textsc{SIGMA}, $r$-band}                                                                                    & arcseconds               \\
\multicolumn{1}{l|}{$R_{\mathrm{e}, \ r}\_\mathrm{err}$}                          & \multicolumn{1}{l|}{Error on effective radius of BCG from \textsc{SIGMA}, $r$-band}                                                                           & arcseconds               \\
\multicolumn{1}{l|}{$n_{r}$}                                            & \multicolumn{1}{l|}{S\'{e}rsic index of BCG from \textsc{SIGMA}, $r$-band}                                                                                  & $-$                      \\
\multicolumn{1}{l|}{$n_{r}\_\mathrm{err}$}                              & \multicolumn{1}{l|}{Error on S\'{e}rsic index of BCG from \textsc{SIGMA}, $r$-band}                                                                         & $-$                      \\
\multicolumn{1}{l|}{$b/a_{r}$}                                          & \multicolumn{1}{l|}{Axis ratio from \textsc{SIGMA}, $r$-band}                                                                                                 & $-$                      \\
\multicolumn{1}{l|}{$b/a_{r}\_\mathrm{err}$}                                          & \multicolumn{1}{l|}{Error on axis ratio from \textsc{SIGMA}, $r$-band}                                                                                                 & $-$                      \\
\multicolumn{1}{l|}{${\theta}_{r}$}                                          & \multicolumn{1}{l|}{Position angle from \textsc{SIGMA}, $r$-band}                                                                                                & degrees                      \\
\multicolumn{1}{l|}{${\theta}_{r}\_\mathrm{err}$}                                          & \multicolumn{1}{l|}{Error on position angle from \textsc{SIGMA}, $r$-band}                                                                                                 & degrees                      \\
\multicolumn{1}{l|}{${{\chi}_{\nu, \ r}}^2$}                                          & \multicolumn{1}{l|}{Reduced ${\chi}^2$, $r$-band}                                                                                                 & $-$                      \\  
\multicolumn{1}{l|}{$m_{i}$}                                        & \multicolumn{1}{l|}{$i$-band apparent magnitude from \textsc{SIGMA}}                                                                                & $-$                      \\
\multicolumn{1}{l|}{$m_{i}\_\mathrm{err}$}                                   & \multicolumn{1}{l|}{Error on $i$-band apparent magnitude from \textsc{SIGMA}}                                                                       & $-$                      \\
\multicolumn{1}{l|}{$R_{\mathrm{e}, \ i}$}                               & \multicolumn{1}{l|}{Effective radius of BCG from \textsc{SIGMA}, $i$-band}                                                                                    & arcseconds               \\
\multicolumn{1}{l|}{$R_{\mathrm{e}, \ i}\_\mathrm{err}$}                          & \multicolumn{1}{l|}{Error on effective radius of BCG from \textsc{SIGMA}, $i$-band}                                                                           & arcseconds               \\
\multicolumn{1}{l|}{$n_{i}$}                                            & \multicolumn{1}{l|}{S\'{e}rsic index of BCG from \textsc{SIGMA}, $i$-band}                                                                                  & $-$                      \\
\multicolumn{1}{l|}{$n_{i}\_\mathrm{err}$}                              & \multicolumn{1}{l|}{Error on S\'{e}rsic index of BCG from \textsc{SIGMA}, $i$-band}                                                                         & $-$                      \\
\multicolumn{1}{l|}{$b/a_{i}$}                                          & \multicolumn{1}{l|}{Axis ratio from \textsc{SIGMA}, $i$-band}                                                                                                 & $-$                      \\
\multicolumn{1}{l|}{$b/a_{i}\_\mathrm{err}$}                                          & \multicolumn{1}{l|}{Error on axis ratio from \textsc{SIGMA}, $i$-band}                                                                                                 & $-$                      \\
\multicolumn{1}{l|}{${\theta}_{i}$}                                          & \multicolumn{1}{l|}{Position angle from \textsc{SIGMA}, $i$-band}                                                                                                & degrees                      \\
\multicolumn{1}{l|}{${\theta}_{i}\_\mathrm{err}$}                                          & \multicolumn{1}{l|}{Error on position angle from \textsc{SIGMA}, $i$-band}                                                                                                 & degrees                      \\
\multicolumn{1}{l|}{${{\chi}_{\nu, \ i}}^2$}                                          & \multicolumn{1}{l|}{Reduced ${\chi}^2$, $i$-band}                                                                                                 & $-$                      \\
\end{tabular}
\label{t:param_list}
\end{minipage}\newline
\label{lastpage}
\end{document}